\documentclass[12pt]{article}
\usepackage{latexsym}
\usepackage{amssymb}
\usepackage{amsfonts}
\usepackage{graphicx}

\catcode`\@=11
\def\numberbysection{\@addtoreset{equation}{section}
        \def\theequation{\thesection.\arabic{equation}}}

\def\be{\begin{equation}}
\def\ee{\end{equation}}
\def\ba{\begin{eqnarray}}
\def\ea{\end{eqnarray}}

\def\und{\underline}
\def\ov{\overline}
\def\I{{\rm Im}}
\def\R{{\rm Re}}
\def\Z{\mathbb{Z}}
\newcommand{\nl}{\nonumber \\}

\def\de{\partial}
\def\a{\alpha}

\def\g{\gamma}

\def\D{\Delta}
\def\d{\delta}
\def\e{\epsilon}
\def\eps{\varepsilon}

\def\th{\theta}

\def\l{\lambda}

\def\m{\mu}

\def\x{\xi}

\def\r{\rho}
\def\s{\sigma}

\def\t{\tau}
\def\f{\varphi}
\def\F{\Phi}
\def\c{\chi}
\def\w{\omega}

\def\wt{\widetilde}
\def\wh{\widehat}

\def\+{{+\!\!\!+}}

\numberbysection

\begin{document}
\begin{titlepage}
\begin{center}
\hfill  \quad LPTHE-03-39 \\
\hfill  \quad DFF 410-12-03 \\

\vspace{1cm} 

{\Large \bf Landau-Ginzburg Description } \\ 
\vspace{.3cm}
{\Large \bf of Boundary Critical Phenomena }\\ 
\vspace{.3cm}
{\Large \bf in Two Dimensions } 

\vspace{1cm}

Andrea CAPPELLI \\
\medskip
{\em I.N.F.N. and Dipartimento di Fisica}\\
{\em  Via G. Sansone 1, 50019 Sesto Fiorentino - Firenze, Italy} \\
\vspace{.5cm}
Giuseppe D'APPOLLONIO,\ \ Maxim ZABZINE\\
\medskip
{\em Lab. de Physique Th\'eorique et Hautes Energies}\\
{\em Universit\'e Pierre et Marie Curie, Paris VI}\\
{\em  4 Place Jussieu, \ 75252 Paris Cedex 05, France}

\end{center}

\vspace{.5cm}
\begin{abstract}
The Virasoro minimal models with boundary are described in the
Landau-Ginzburg theory by introducing
a boundary potential, function of the boundary field value.
The ground state field configurations become non-trivial and are found to
obey the soliton equations. 
The conformal invariant boundary conditions are characterized by
the reparametrization-invariant data of the boundary potential, 
that are the number and degeneracies of the stationary points.
The boundary renormalization group flows are obtained by varying
the boundary potential while keeping the bulk critical:
they satisfy new selection rules and correspond to real 
deformations of the Arnold simple singularities of $A_k$ type.
The description of conformal boundary conditions in terms of boundary
potential and associated ground state solitons is extended to the
$N=2$ supersymmetric case, finding agreement with the analysis
of $A$-type boundaries by Hori, Iqbal and Vafa. 
\end{abstract}

\vfill
\end{titlepage}
\pagenumbering{arabic}


\section{Introduction}

Large classes of conformal invariant boundary conditions have been 
found in rational conformal field theories over the recent years  
by developing several algebraic methods. 
The simplest, fully symmetry-preserving conformal boundary states are
the so-called Cardy states,
that exist for left-right symmetric sectors in the bulk \cite{cardy}.
Conformal boundaries have also been found 
in the case of non-diagonal bulk sectors or
in presence of partial symmetry breaking at the boundary
(still preserving conformal invariance) \cite{comcon}.
For many theories, like the Virasoro minimal models, the conformal
boundary states have been completely classified \cite{zuber}.

Far less is known about non-conformal boundary conditions, where a relevant
interaction breaks scale invariance at the boundary and
leads to a renormalization group (RG) flow toward another conformal
boundary of the same bulk critical theory.
Two relevant cases have been extensively studied during the last decade:
the Kondo problem of magnetic impurities in a diamagnetic metal \cite{kondo}
and the resonant scattering of edge excitations in the quantum Hall effect
 \cite{hall}.
The main present motivation for studying boundary RG flows comes from 
open string theory, where the decay of an
unstable brane (or of a stack of branes) can be modeled by
boundary interactions in the corresponding world-sheet conformal theories
-- the so-called ``tachyon condensation'' \cite{tachyon}.

Although the boundary renormalization group flows are formally and 
perturbatively analogous to the bulk flows, they
may have specific features, and it is useful to have 
physical descriptions of the boundary dynamics, such as that of Ref. 
\cite{saleur}.
In the case of the bulk RG flows, the Landau-Ginzburg  mean-field theory
has been a valuable source of intuition:
here, the multicritical points are obtained by fine-tuning 
the parameters in the scalar potential, and the flows 
to lower critical points follow by detuning.
This description yields a qualitative chart of the space of flows, 
basically its topology, that survives quantum corrections
even in two dimensions \cite{zamlg}.

Furthermore, the Landau-Ginzburg (LG) theory becomes exact in presence of
N=2 supersymmetry, due to the non-renormalization theorems. 
The beautiful works by Martinec, Vafa, Warner and others \cite{vw}
have shown that the study of the multicritical points
of Landau-Ginzburg potentials can be mapped into 
Arnold's classification of singularities of analytic functions modulo
reparametrizations \cite{arnold}.
This relation also provides a physical explanation to the fact
that the Virasoro minimal models, the simpler multicritical points 
in two dimensions, are classified by A-D-E Dynkin diagrams, as well as 
the Arnold simple singularities \cite{ciz}.

The present understanding of boundary multicriticality and boundary
RG flows is based on the study of integrable systems, both
in the continuum and on the lattice, and on numerical analyses. 
For many theories
with critical points of known conformal type, integrability has been
extended in the presence of boundary conditions and boundary interactions.
For example, lattice statistical models are known to realize all the 
boundary conditions of the Virasoro minimal models \cite{pearce} 
and examples of
boundary integrable RG flows have been obtained by specializing 
integrable bulk interactions to the boundary \cite{saleur}.
Known numerical methods, such as the truncated conformal space approach 
\cite{watts}
and the thermodynamic Bethe ansatz \cite{tba}, have been applied to 
the study of boundary interactions.
In some approaches, such as integrable scattering theory at the
boundary, the boundary dynamics is described by 
quantum mechanical degrees of freedom that
must be added at the boundary for completeness \cite{saleur}.
Upon integrating them out, one may generate complicate non-local
interactions at the boundary.

In this paper, we describe the generalization of the Landau-Ginzburg
 theory in presence of a boundary. We do not introduce new
degrees of freedom at the boundary, but rather describe the
dynamics in terms of a tunable scalar boundary potential 
$V_b(\f_o)$, function of the boundary value $\f_o$ of the scalar field 
$\f$, whereby implicitly assuming locality\footnote{
This is the traditional approach of LG studies in $(4-\e)$ dimensions
\cite {binder}.}.
Using these rather restrictive assumptions, but consistent with
the results of the corresponding lattice models \cite{pearce}, we obtain a 
comprehensive description of 
the boundary conditions and RG flows in the Virasoro models of
the so-called (A,A)-series. 
The outcome is a rather simple extension of the LG description
of multicriticality by fine-tuning, 
showing again deep relations with Arnold's work.
Our analysis also applies to
the corresponding $N=2$ supersymmetric minimal models with $A$-type
boundary conditions, where it matches
the earlier inspiring results of Ref.\cite{vafa}.

In Section two we introduce the LG theory with boundary: we consider
the $m$-th Virasoro minimal model in the (A,A) series, that
corresponds to a $(m-1)$-fold critical generalization of the Ising model
and is described by the LG theory with bulk potential $V=\l\f^{2(m-1)}$ 
\cite{zamlg}.
The allowed boundary potentials turn out to be of the form
$V_b=a\f_o^{m-2}+b\f_o^{m-3}+ \dots\ $, 
corresponding to the real deformations of   
the Arnold singularity $x^m$, associated to the $A_{m-1}$ Dynkin diagram 
\cite{arnold}. 
We study the non-trivial ground state solutions generated by the boundary term
and show that they are  related to the solitons of the bulk potential.
At bulk criticality, we divide the solutions into equivalence classes
modulo field reparametrizations at the boundary, and argue that each class
models (i.e. renormalizes to) one conformal invariant boundary
condition, whose properties match the known data from conformal field
theory \cite{cardy}\cite{zuber} and integrable lattice models \cite{pearce}.
We then find that each conformal boundary condition can be associated to
a sub-diagram of the $A_{m-1}$ Dynkin diagram, that specifies
the type of stationary point of the boundary potential.
In this description, the non-degenerate solutions of the boundary
conditions match the stable conformal boundaries, while the
degenerate solutions correspond to unstable boundaries, possessing
a number of relevant boundary fields equal to the order of
degeneracy.
The extension of this analysis to the (A,D) and (A,E) series 
(Potts model and generalization and exceptional cases), presents some
limitations inherited from the bulk LG theory \cite{kekeli}: these issues 
are discussed in the Appendix A, using the Potts model as an example.

In Section three, the boundary RG flows are described by detuning 
the boundary potential out
of one degenerate stationary point to produce less degenerate points ---
the same mechanism of bulk RG flows working in a richer
nested pattern.
We then find new selection rules for the RG flows out of a Cardy boundary,
that are simply obtained by breaking the associated Dynkin diagram
into small diagrams representing the boundaries reached in the infrared
limit; these rules confirm and extend the results of Ref.\cite{rrs,kevin}.
Moreover, the topology of the space of RG flows 
can be deduced from the parameter space of real deformations of
Arnold singularities. The analysis of the tri- and tetra-critical
Ising models are worked out explicitly, finding agreement 
with earlier results from integrable models \cite{ti-num}
and perturbative calculations \cite{watts}.

In Section four, we extend the analysis of ground state solutions 
with non-trivial boundary to the LG theory with N=2 supersymmetry \cite{vw}; 
earlier descriptions of the $A$-type supersymmetric boundary 
conditions had already established a connection with soliton equations
and singularity theory \cite{vafa}.
The supersymmetric boundary conditions are again 
determined by the form of the boundary scalar potential and have a
regular description at bulk criticality, namely they fit
the picture established in the non-supersymmetric case. 
However, the detailed aspects of the
boundary solutions are different in the supersymmetric case
and their pattern of RG flows remains to be understood.


\subsection{Boundary conditions in the Virasoro minimal models}

We start by recalling the properties of the Ising and Tricritical Ising 
models, that will be used for introducing the main points of our approach.
The bulk conformal theory of the Ising model
contains three left-right symmetric sectors and, correspondingly, there are
three Cardy conformal boundary conditions \cite{cardy}: 
they are called $(+)$, $(-)$,
representing fixed spin values, and $(f)$ for the free conditions.
The corresponding index pairs in the Kac table are, $(r,s)=(1,1)$, $(1,2)$
and $(2,2)$, respectively.
By switching on a boundary magnetic field $H_b$, one obtains a RG flow
between the free and the fixed boundary conditions.
Therefore, $(\pm)$ are stable boundary conditions and $(f)$ is once unstable, 
namely there is one relevant boundary field compatible with
the boundary condition. The allowed boundary fields are listed in the
partition function on the upper-half plane, that is conformally
equivalent to that of the strip with equal boundaries on the 
two sides\footnote{
We follow the notations of Ref.\cite{zuber}.} \cite{cardy},
\be
Z_{a|a}= \sum_{b}\ n_{aa}^b\ \c_b\ ,\qquad\qquad 
a\equiv (r,s), b\equiv (r',s'),
\label{stripz}
\ee
where $n_{aa}^b$ are the fusion coefficients and $\c_{r,s}$ the Virasoro
characters. In particular,
$Z_{f|f}=\c_{1,1}+\c_{1,3}$.

\begin{figure}[ht]
\begin{center}
\includegraphics[width=12cm]{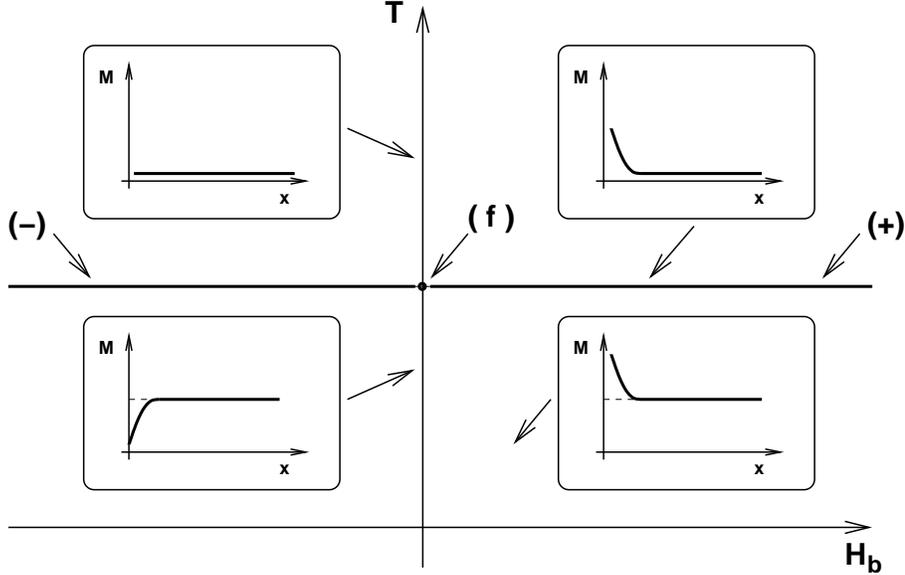}
\end{center}
\caption{Phase diagram of the Ising model with boundary in two dimension:
$T$ is the bulk temperature and $H_b$ the boundary magnetic field.
Magnetization profiles $M(x)$ are drawn in the insets, with 
$x$ measuring the distance from the boundary.}
\label{fig-ising}
\end{figure}

The phase diagram is shown in Figure \ref{fig-ising}.
We also sketch the magnetization profile in all the
regions: in the broken phase, the magnetization vanishes at the
free boundary, while it is enhanced by a boundary magnetic field. 
Two facts are relevant for the following discussion:

i) In more than two dimensions, there is spontaneous magnetization at
the boundary, caused by another parity-invariant, relevant boundary 
parameter, the boundary spin coupling, and there is an associated 
boundary tri-critical point \cite{binder,henkel}.
The two-dimensional boundary phase diagram is much simpler, because
spontaneous magnetization cannot occur on the one-dimensional boundary
(for $H_b <\infty$).

ii) At bulk criticality, $T=T_c$, the free boundary condition corresponds to
a vanishing boundary field value, that is manifestly conformal invariant.
The boundary conditions for $H_b \neq 0$ instead correspond to non-vanishing
boundary values that are not invariant. 
Here the $(\pm)$ conformal boundary conditions 
can be identified as follows:
one should consider the continuum limit $\D x \to 0$ of the Ising model
near  $T=T_c$ and in the presence of the boundary condition.
Since the renormalized spin field $\s_R$ acquires a positive scaling
dimension $\d$, any non-vanishing boundary spin value $\s_o$
renormalizes to infinity, $\s_R =\s_o/\D x^\d \to \pm\infty$:
the conformal boundary conditions $(\pm)$ are in fact $(\pm\infty)$ 
\cite{cardy2}. 

The Tricritical Ising model is another well-studied example: 
the tricritical bulk point is described by the $m=4$ Virasoro
minimal model, having six left-right symmetric bulk sectors.
The corresponding six conformal boundary conditions have been interpreted
in the spin model and their boundary RG flows have been found
by using the results of integrable models \cite{chim}, numerical analyses 
\cite{watts} and perturbative calculations \cite{rrs}.
In the lattice description of an Ising model with vacancies,
there naturally are three stable boundary conditions, corresponding to
fixed spin values:
\be
(+)=(1,1),\qquad (-)=(3,1), \qquad(0)=(2,1) \ \ {\rm (no\ spin)}.
\label{ising-fix}
\ee 
Next there are the boundaries $(0+)$, $(0-)$, corresponding to 
partially fixing the spins on the boundary, 
and finally the free conditions $(f)$ (also called $(d)$ for
``degenerate''):
\be
(0+)=(1,2),\qquad (0-)=(1,3), \qquad(f)=(2,2) \ .
\label{ising-bc}
\ee 
We report their partition functions for later use:
\ba
Z_{(0+)|(0+)} &=& Z_{(0-)|(0-)} = \c_{1,1}+\c_{1,3}\ ,\nl
Z_{(f)|(f)} &=&  \c_{1,1}+\c_{1,2}+ \c_{1,3}+\c_{1,4}\ .
\label{tricriz}
\ea
The fields $\phi_{1,2}$ and $\phi_{1,3}$ are relevant,
so the boundaries $(0\pm)$ and $(f)$ are once and twice unstable,
respectively.

\begin{figure}[ht]
\begin{center}
\input{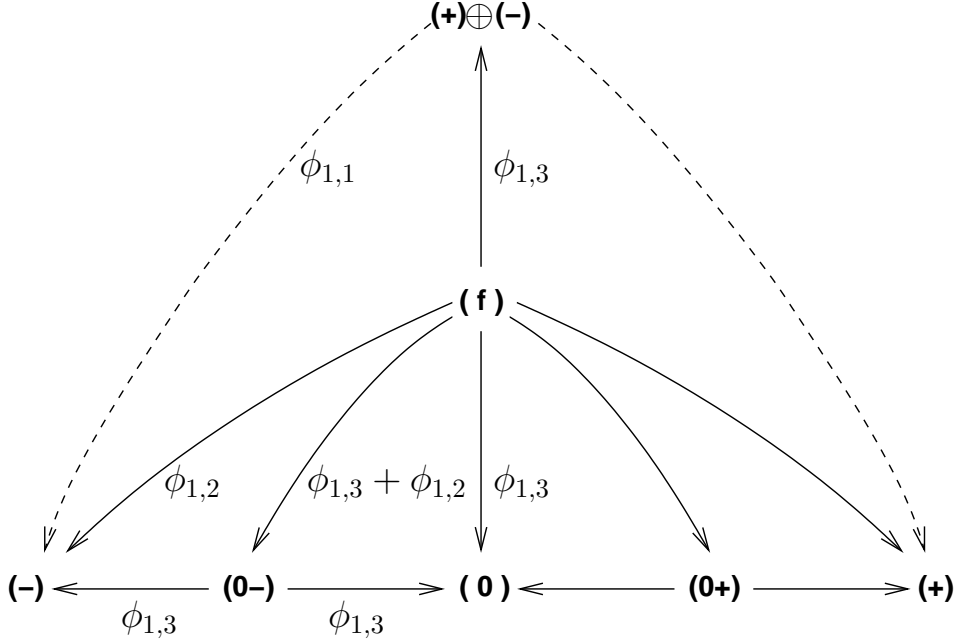}
\end{center}
\caption{Space of boundary RG flows in the Tricritical Ising model
from Ref.\cite{affleck}: the dashed lines represent first-order 
transitions; the driving field is reported for each flow.}
\label{fig-tricri}
\end{figure}

The space of RG flows  is reported in Figure \ref{fig-tricri} \cite{affleck}.
There are two relevant directions out of $(f)$, 
one is $\Z_2$ even ($y$ axis, $\phi_{1,3}$ field) and the other 
one is odd ($x$ axis, $\phi_{1,2}$ field).
All flows verify the conjectured ``$g$-theorem'' that establishes
that the boundary entropy $g$ should decrease along the flow \cite{g-th}.
A novel feature, first observed perturbatively in Ref.\cite{rrs}, is that
one can flow from a single Cardy state, $(f)$, to a superposition
of them, $(+) \oplus (-)$. The latter boundary 
has one relevant perturbation, given by
the second identity field: the superposition can be disentangled 
by applying a boundary magnetic field leading to a first-order phase transition
at the boundary (namely a discontinuous jump in the boundary magnetization)
\cite{affleck}.

The boundary dynamics in the higher Virasoro models
is only partly known. The critical points in the bulk follow the so-called
A-D-E classification, namely each model can be associated to a pair of
simply-laced Dynkin diagrams (A,G), where G can be A, D or E, such that
one has two models for each value of $c(m)=1-6/m(m+1)$, $m=5,6,\dots$, 
of type (A,A) and (A,D), and a third one of type (A,E) for some special
$m$ values \cite{ciz}.

We shall mostly deal with the (A,A) series of multicritical 
Ising models, that can be realized by the 
solid-on-solid integrable lattice models \cite{bauer}.
The bulk conformal theory possesses left-right diagonal partition function
and the properties of the conformal boundaries can be easily obtained 
from Cardy's analysis \cite{cardy}. 
The partition functions on the
upper half plane $Z_{(r,s)|(r,s)}$, is of the form Eq.(\ref{stripz}), with
indexes $1\le r \le m-1$, $1\le s \le m$, and $(r,s)\sim (m-r, m+1-s)$.

\begin{figure}[ht]
\begin{center}
\includegraphics[width=6cm]{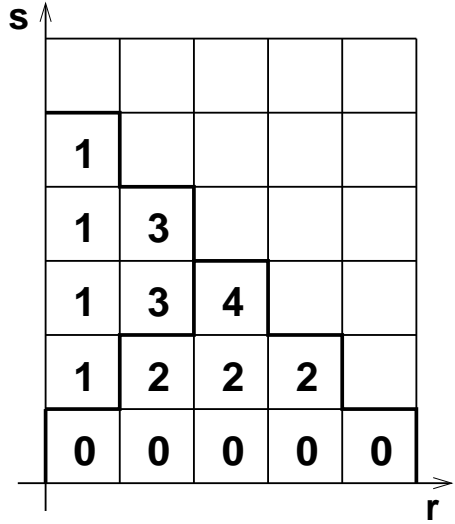}
\end{center}
\caption{Number of relevant fields for each boundary condition of
Kac's indexes $(r,s)$ for $m=6$.}
\label{fig-aztec}
\end{figure}

We can read the boundary operator content from the partition function
and count the number of relevant fields for each boundary.
In Figure \ref{fig-aztec}, we report this number in the corresponding 
$(r,s)$ box of the Kac table for the $m=6$ model (penta-critical Ising).
One sees the pattern of two ``Aztec pyramids'', with even (resp. odd)
number of relevant fields: there are $(m-1)$  stable boundaries, with indexes
$(r,1)$,  $1\le r \le m-1$, then $(m-2)$ once unstable boundaries 
$(1,s)$, $2\le s \le m-1$, next $(m-3)$ twice unstable ones and so forth,
up to a total of $m(m-1)/2$ conformal boundaries.

These boundary conditions have been described in the lattice model
as follows \cite{pearce}.
The bulk spin or ``height'' configurations can take $m$ values,
say $h=1,\dots,m$, jumping by $\pm 1$ on neighbor sites, according
to the adjacency rule of the points on the $A_m$ Dynkin diagram; 
thus, there are $(m-1)$ bulk phases of fixed magnetization 
and a corresponding number of fixed boundary conditions.
The other, unstable conformal boundaries have also been uniquely identified
in the lattice model by checking the fusion
of pairs of boundaries \cite{defects}.
The corresponding lattice conditions partially fix the spin at the boundary,
that can span increasingly larger ranges of values, as 
given by the ``fusion'' of adjacent points in the Dynkin diagram \cite{pearce}.
In conclusion, all boundary conditions are local in the spin variable
and, moreover, they
allow fluctuations gradually raising from nothing (fixed) to full (free): 
this suggests the association of the degree of boundary fluctuations with
that of RG instability, a fact that will be later implemented in 
the LG description.

Another interesting property of the (A,A) models is the $\Z_2$ spin parity,
that acts on the boundary states as follows: 
\be
{\cal P} :\ (r,s)\ \longrightarrow\ (m-r,s)\sim (r,m+1-s)\ ,
\label{z2}
\ee
such that the boundaries can be classified in singlets and doublets: 
moreover, boundary fields have a definite parity if 
they appear on invariant boundaries, otherwise they do not.
Let us report the results of the analysis in Ref.\cite{ruelle}.

For $m$ odd, the invariant boundaries are,
$a=(r,(m+1)/2)$, $r=1,\dots,(m-1)/2$, and the sign of the fields
that can appear in $Z_{a|a}$ is:
\be
{\cal P} :\ \phi_{r,s}\ \longrightarrow\ 
(-1)^{(r+1)\left(\frac{m+1}{2}+1 \right)+\frac{s-1}{2}}
\ \phi_{r,s}\ , \qquad s\ {\rm odd}.
\label{z2odd}
\ee
For $m$ even, the invariant boundaries are,
$a=(m/2,s)$, $s=1,\dots,m/2$, and the sign of the fields is:
\be
{\cal P} :\ \phi_{r,s}\ \longrightarrow\ 
(-1)^{(s+1)\left(\frac{m}{2}+1 \right)+\frac{r-1}{2}}
\ \phi_{r,s}\ , \qquad r\ {\rm odd}.
\label{z2even}
\ee
Note finally that boundaries related by the $\mathbb{Z}_2$ symmetry have 
equal boundary entropy, $g_{(r,s)}=g_{(m-r,s)}=g_{(r,m+1-s)}$.

\section{Landau-Ginzburg theory with boundary}


\subsection{Bulk critical theories and bulk renormalization group flows}

Let us start by recalling the LG description of the (A,A) Virasoro
minimal models in the boundaryless case \cite{zamlg}. 
We consider a real scalar field with action:
\be
S=\int d^2 x \ \frac{1}{2}\left( \partial_\mu \f \right)^2
+ V\left(\f\right) \ .
\label{LGbulk}
\ee
The $m$-th minimal model corresponds to the multicritical point of
$(m-1)$ coexisting phases; therefore, the $\Z_2$ symmetric potential
should have $(m-1)$ minima merging at the critical point, namely:
\be
 V\left(\f\right) = V_{\rm crit}\left(\f\right) \equiv 
\l \ \f^{2(m-1)} \ , \qquad\qquad
{\rm (at\ criticality)}.
\label{LGbulkV}
\ee
The coupling $\l$ is relevant near the ultra-violet (UV) point
and drives the scalar theory to the interacting infra-red (IR) fixed point
corresponding to the minimal model; at this point, $\l$ becomes marginal.

Further RG flows to  lower-critical models are described as follows:
the theory (\ref{LGbulk},\ref{LGbulkV}) at the IR point still possesses 
the relevant field $\f^n$, $n=1,\dots, 2(m-2)$, corresponding to detuning
the potential (\ref{LGbulkV}) out of the multicritical point
(the field $\f^{2m-3}$ is redundant due to the equation of motion). 
The flow to the $k$-th minimal model, $k<m$,  is obtained by
adding the term $\f^{2(k-1)}$ to the potential:
\be
V_{UV}=\l \ \f^{2(m-1)} + \eta\ \f^{2(k-1)} \quad\longrightarrow\quad
V_{IR}= \eta\ \f^{2(k-1)} \ , 
\label{LGbulkRG}
\ee
At the $k$-th infrared fixed point, the coupling $\eta$ becomes itself
marginal, while the original coupling $\l$ becomes irrelevant and 
should be discarded\footnote{
This is the case $\eta>0$; for $\eta<0$ one clearly flows to a completely
massive phase.}.

Although this classical theory gets very strong quantum corrections
in two dimensions, the qualitative picture remains valid and
it describes the chain of multicritical points in
the Virasoro minimal models.
It also explains the irreversibility of the RG flow in terms 
of the detuning of a highly degenerate minimum of the potential.
Further evidence for the LG theory is
found by matching the relevant fields in the two
theories, and by analyzing their symmetries, fusion rules and RG flows,
as nicely shown in the Refs.\cite{zamlg}.


\subsection{Landau-Ginzburg theory with boundary and soliton equations}

We now consider the theory on the Euclidean half-plane with coordinates
$x \ge 0$ and $\t$, and add a boundary potential $V_b$ to the Lagrangian 
as follows,
\be
S=\int_{x>0} d^2 x \ \frac{1}{2}\left( \partial_\mu \f \right)^2
+ V\left(\f\right) \ + \ \int dt\ V_b\left(\f_o\right)\ ,\qquad\qquad
\f_o\equiv \f(x=0)\ .
\label{LGaction}
\ee
This is actually the traditional approach in $(4-\e)$ dimensions
\cite{binder}. We assume that the bulk potential $V$ 
possesses up to $(m-1)$ distinct minima $\f=v_i$, $i=1,,\dots,m-1$,
at the same height, $V(v_i)=0$, since we are interested in the bulk
multicritical point and the nearby region of coexisting phases.
The ground state field configuration is obtained by extremizing the action:
we consider a static, finite-energy solution, with 
the field approaching a minimum of $V(\f)$ in the bulk,
$\lim_{x\to\infty}\f(x) = \f_\infty=v_i$, with vanishing derivative.
The stationary conditions read:
\ba
0&=&\frac{\partial^2\f}{\partial x^2} - \frac{\partial V}{\partial\f} \ ,
\label{LGeq}\\
0&=&\d\f_o\left(\left.
\frac{\partial\f}{\partial x}\right\vert_0 - 
\frac{\partial V_b}{\partial\f_o} 
\right)\ .
\label{LGbc}
\ea
We recall that the variational principle let
us choose whether to vary the field at the spatial boundary or not:
since the (target) field space is represented by the real line,
we may consider boundary conditions corresponding to D0 branes, i.e. 
given points on the line, or to one D1 brane corresponding
to the complete line.
In the first case, we do not vary the boundary field, 
$\d\f_o=0$ and get no conditions
from the boundary potential: the solutions are $(m-1)$ constant
field profiles fixed at the bulk minima, $\f\equiv v_i$.
These are stable boundary conditions by definition.

In the case of D1 branes, we do vary the boundary field and obtain a 
boundary condition depending on the tunable $V_b$. This is the 
interesting and generic situation, since the D0 branes will be
recovered from localization by the boundary potential.
The equation of motion (\ref{LGeq}) can be solved following the well-known
analysis of the solitons in the multi-valley $V(\f)$: it describes
a classical particle moving in the upside down 
potential $U=-V$, having at least
two maxima at the same height, $V(v_1)=V(v_2)=0$.
The particle start at ``time'' $x=\infty$ from one of the bulk minima
with vanishing velocity and energy, and reaches the 
boundary $\f_o$, with velocity determined by the boundary condition
(\ref{LGeq}). 
This non-trivial field configuration corresponds to the classical
ground state of the system in the presence of the boundary, and should be
distinguished from the usual solitons in the bulk that correspond
to excited states.

Let us first consider the simplest case of the Ising model, 
$m=3$, with just two minima $v_1=-v_2=v$ ($\l=1/2$):
\be
V=\frac{1}{2}\ \left(\f^2-v^2\right)^2 \ ,\qquad\qquad\qquad 
{(\rm Ising\ model)}.
\label{Vising}
\ee

\begin{figure}[ht]
\begin{center}
\input{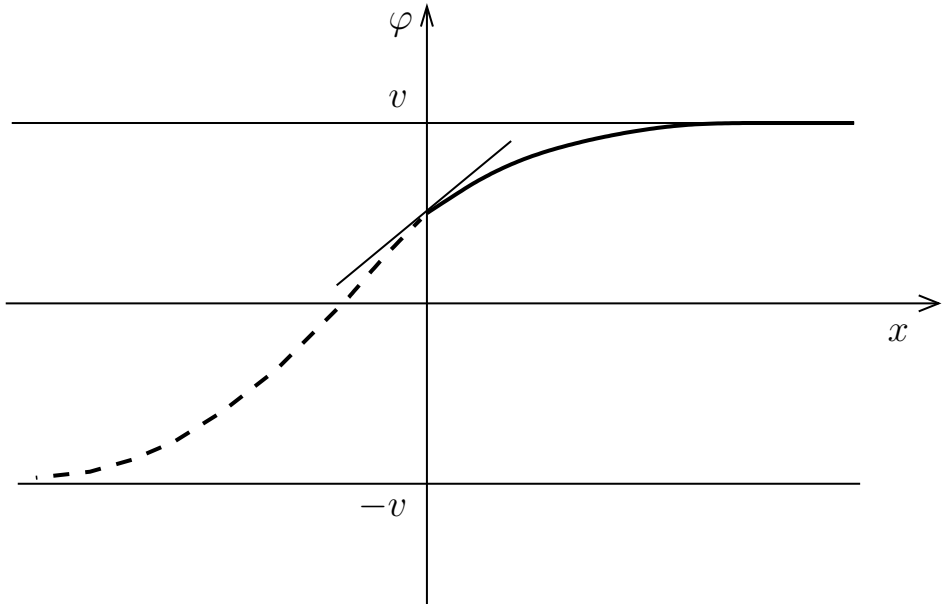}
\end{center}
\caption{Ground state field profile corresponding to a cut soliton.}
\label{fig-soli}
\end{figure}

The ground state field profile is given by a soliton
of the bulk theory cut at some point (Figure \ref{fig-soli}): 
we may have the trivial solitons, $\f=\pm v$, the usual soliton,
$\f =\pm v\ \tanh\left(v(x-\x)\right)$, and the singular solitons
involving the cotangent.
We can also consider superpositions of ground state solutions if degenerate
in energy.

Our task here is to represent the conformal-invariant boundary
conditions at bulk criticality in terms of carefully chosen
solitons solutions for $v\to 0$ (cf. Figure \ref{fig-ising}).
It is natural to divide them into ``universality classes''
whose characteristic proprieties survive the limit $v\to 0$.
Part of the motivations for carrying out this program are coming
from the analysis of the N=2 supersymmetric case \cite{vafa},
where it was shown that the $A$-type supersymmetric boundary
conditions are in one-to-one relation with the solitons of the bulk theory
and that these form universality classes whose critical limits match
the (supersymmetric) conformal boundaries.

We thus formulate the following hypothesis:
\ba
{conformal\ boundary}&\sim&
{universality\ class\ of\ ground\ state\ profiles} \nl
&\sim& {universality\ class\ of\ soliton\ solutions}
\ea
The equation of motion (\ref{LGeq}) can be integrated once to obtain the
conservation of the particle energy,
\be
\frac{\partial\f}{\partial x}=\pm\ \sqrt{2\left( 
V\left(\f\right)-V\left(\f_\infty\right) \right)} \ ,
\label{sol-eq}
\ee
that determines the field profile by further integration.
Upon evaluating the energy at the boundary, we get another boundary
condition that can be combined with (\ref{LGbc}):
\be
\left.\frac{\partial\f}{\partial x}\right\vert_0 \ =\  
\frac{\partial V_b}{\partial\f_o} \ =\ \pm\ \sqrt{2\left( 
V\left(\f_o\right)-V\left(\f_\infty\right) \right)} \ .
\label{full-bc-eq}
\ee
The second and third terms give an algebraic equation 
for the boundary value $\f_o$, that will be very important in
the following; this in turns determines the boundary
derivative (the first term in (\ref{full-bc-eq})), 
since the equation of motion
has one boundary condition left to be set.

One important property of the classes of soliton solutions is 
given by their asymptotic value in the bulk.
We shall disregard boundary phases different from that of the bulk 
(boundary interfaces or ``surface wetting'' \cite{binder}) that can be
presumably discussed by tensoring boundaries \cite{defects}.
Moreover, the absence of spontaneous magnetization
at the one-dimensional boundary suggests us to discard
solutions with boundary critical points of
higher order than that in the bulk.

In the Ising case, we can identify two classes of solitons, one for
each asymptote $\f_\infty=\pm v$, having $\f_o$ a fixed non-vanishing
value of the same sign of $v$ (cf. Figure \ref{fig-ising});
these solitons solve Eq. (\ref{full-bc-eq}) with an appropriate
value of $V_b=-a\f_0$, and naturally have $v\to 0$ limit in the two fixed
conformal boundaries $(\pm)$
(keeping in mind that the scaling limit would naturally drive
$\f_o\to\infty$).
The two type of solitons exchange under parity as the conformal boundaries do.
Each type exists in the respective phase, and they
are simultaneously present at phase coexistence close to bulk criticality.
They describe stable boundaries because small variations of $\f_o$
do not change the class of solutions.

As for the free conformal condition $(f)$, we should consider
the tanh soliton with $\f_o=0$; there are two of them, $\f_\pm(x)$
approaching $\f_\infty=\pm v$. Since the $(f)$ boundary is parity invariant, 
we should consider a superposition of the two solutions,
of the form $Z=Z[\f_+(x)]+Z[\f_-(x)]$, where 
$Z[\f] = \exp\left(-S_{{\rm cl}}[\f]\right)$.
Next, we remark that one parity-invariant solution is possible
at bulk criticality, $v=0$, in the case of $V_b=0$:
\be
\left.\frac{\partial\f}{\partial x}\right\vert_0 \ =\ 0 =
\ \pm\ \f_o^2 \ .
\label{bc-eq0}
\ee
This solution is degenerate and can be thought of
as the $v\to 0$ limit of the previous
superposition $Z=Z[\f_+(x)]+Z[\f_-(x)]$.
The instability of this boundary condition is related to
its degeneracy, since any addition of $V_b\neq 0$
would lead to a non-degenerate solution $\f_o\neq 0$ breaking
the parity invariance (and altering the superposition 
$Z=Z[\f_+(x)]+Z[\f_-(x)]$ for $v\neq 0$).
Indeed, the boundary equations (\ref{full-bc-eq}) with a non-vanishing
boundary magnetic field have the solutions, for $v=0$:
\be
V_b\ =\ -H_b\ \f_o\ \ \longrightarrow \left\{
\begin{array}{lll}
\f_o\ = \sqrt{|H_b|} && H_b>0\ ,\\ 
\f_o\ = -\sqrt{|H_b|} && H_b<0\ .    \end{array}
\right.
\label{ising-RG}
\ee
The two solutions are representative of the $(\pm)$ conformal conditions,
as said before.

In conclusion, we have described the 
three Ising conformal boundaries in terms of classes of 
ground state profiles at bulk criticality (and in the neighbors of it), 
in agreement with the known facts and the expected boundary RG flows.
We remark that in more than two dimensions
the $\Z_2$ even term, $V_b=a\ \f_o^2$, can be added to the boundary 
potential \cite{binder}, causing spontaneous
magnetization at the boundary, i.e. simultaneous solutions $\f_o=0$ and 
$\f_o\neq 0$ of the algebraic equation (\ref{full-bc-eq});
this term should be discarded in two dimensions and indeed it will
be found to be redundant.


\subsection{Conformal boundary conditions and Arnold's singularities}

We can now formulate the general strategy for finding the classes
of solutions matching the Virasoro conformal boundaries:
\begin{itemize}
\item Take the critical limit in the bulk, such that all ground state 
solutions are simultaneously present in the LG equations.
\item Study the solutions of the algebraic equation for $\f_o$ 
(\ref{full-bc-eq}),
that can be rewritten as the stationary condition of the ``superpotential''
$W$:
\ba
 \frac{\partial W}{\partial \f_o} &=& 0\ , \nl
W &=& \mp \int^{\f_o}_0 d\f \sqrt{2V_{crit}}\ +\ V_b\nl
&=& \mp \frac{\f_o^m}{m}\ +\ 
a_{m-3} \frac{\f_o^{m-2}}{m-2} \ +\ a_{m-4}\frac{\f_o^{m-3}}{m-3}\ +\ 
\cdots \ +\ a_0\ \f_o\ . \nl 
\label{alg-eq}
\ea
\item Identify stationary points of $W$ up to field reparametrizations
at the boundary $\f_o \to \f_o + \e (\f_o)$.
\item Map stationary points to conformal boundaries as follows:
\ba
\!\!\!\!\!\!\!\!\!\!\!\!\!\!
non-degenerate\ stationary\ points\ of\ W  &\leftrightarrow &
stable\ boundaries \nl
\!\!\!\!\!\!\!\!\!\!\!\!\!\!
n\!-\!fold\ degenerate\ stationary\ points\ of\ W & \leftrightarrow &
n\!-\!fold\ unstable\ boundaries \nl
\label{main-hp}
\ea
\end{itemize}
In this approach, the boundary multicriticality is recast 
into a form similar to the
bulk case, where the RG flows out of a conformal boundary is
associated to the detuning of a degenerate critical point \cite{zamlg}.
The hypothesis of field reparametrization invariance 
is similarly justified (``universality'') and it leads to
the pattern of the Virasoro boundaries described in Section 1.1,
as we now discuss.

The stationary points of $W(\f_o)$ can be analyzed by using
some results of the Arnold theory of singularities
(of the inverse map), here in the real domain  \cite{arnold}.
The deformations of the $w=x^m$ singularity are described by adding the
polynomials $p(x)$, $w\to w+p(x)$, and should be identified
modulo reparametrizations $x\to x+ q(x)$; they form the ring,
$\mathbb{Q}=p(x)/(q(x)dw/dx)=\{1,x,x^2,\dots,x^{m-2} \}$
of dimension $(m-1)$ \cite{vw}.
This matches the number of relevant fields
of the most unstable Virasoro boundary (see Section 1.1),
plus one for the identity field. 
Thus, we associate this boundary with the most degenerate 
stationary point of (\ref{alg-eq}) for $V_b=0$.
Note that the redundant perturbation $x^{m-1}$ is discarded
(as in the Ising case), and there are no boundary states
of higher criticality than that in the bulk.

The other conformal boundaries are related to the less degenerate 
stationary points.  
The polynomial $w=x^m+p(x)$ has at most $m-1$ such points: let
us first take them to be all distinct on the real line, where
they can be ordered by numbering them from $1$ to $m-1$.
These points can be moved upon varying the parameters in $w$;
a pair of neighbor points can meet to form a degenerate stationary point, and 
further deformations move them into the complex
plane, i.e. they disappear in pairs; there are no exchanges of
singularities in the real case, thus ordering always hold.
Therefore, we can count, besides the $m-1$ simple stationary points,
$m-2$ once degenerate ones, $m-3$ two-fold degenerate ones
and so forth, till the unique $(m-2)$-fold degenerate point.
This is precisely the pattern of RG instability of the Virasoro
conformal boundaries (Figure \ref{fig-aztec}), because
the $(n-1)$-fold degenerate point (merging of $n$ stationary points)
has $n-1$ unfolding parameters --- the relative distances
between the points.
Note that this description of the stationary points on the real line is
stable under smooth reparametrizations of $x$, that
can move the points but cannot change the order of the singularities.

\begin{figure}[ht]
\begin{center}
\includegraphics[width=6cm]{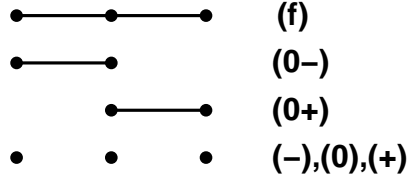}
\end{center}
\caption{$A_{m-1}$ Dynkin diagram and associated subdiagrams for
$m=4$, the Tricritical Ising model. The labels of the conformal boundaries
are reported next to the diagrams.}
\label{fig-dynkin}
\end{figure}

The pattern of stationary points can be visualized by using Dynkin diagrams.
In Arnold's theory, the singularity $w=x^m$ is associated to the
$A_{m-1}$ diagram (Figure \ref{fig-dynkin}): the ordered $m-1$ points
can be drawn on the real line representing the non-degenerate
stationary points of $w$; two points joined by a segment can
represent a once degenerate point, further joining gives the higher
degenerate points, up to the
full diagram representing the highest degenerate point.

\begin{figure}[ht]
\begin{center}
\input{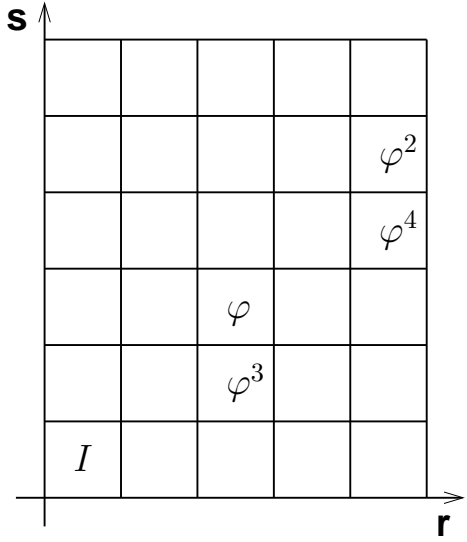}
\end{center}
\caption{LG description of the relevant fields for the most unstable boundary 
$(r,s)=(3,3)$ of the $m=6$ model.}
\label{fig-fields}
\end{figure}

Further support for this  description of Virasoro
boundaries is provided by the identification of the relevant fields
of the most unstable boundary and their parity symmetry.
Let us take the $m=6$ case for example (Figure \ref{fig-aztec}).
The most unstable boundary state $(r,s)=(3,3)$ is represented here
by the most degenerate solution of the stationary equation (\ref{alg-eq}),
namely $\f_o^5=0$ for $V_b=0$.
The deformations of this singularity are parametrized by
the terms in the boundary potential $V_b$: $\f_0, \f_0^2,
\f_0^3,\f_0^4$.
On the other hand, the relevant boundary conformal fields
are listed in the partition function $Z_{(3,3)|(3,3)}$ (\ref{stripz}):
they are on the first two diagonals of the Kac table, 
one every second field: $\phi_{(3,3)}$, $\phi_{(5,5)}$, and 
$\phi_{(3,2)}$, $\phi_{(5,4)}$. The matching of the fields in the
two descriptions is shown in Figure \ref{fig-fields}.
The $\Z_2$ symmetry of the conformal fields was discussed in Section 1.1, 
Eqs. (\ref{z2odd},\ref{z2even}), and is found to agree with that 
of corresponding LG field powers.

In the $m$-th minimal model, the identification of the relevant fields
for the most unstable boundary state, $(r,s)=(m/2,m/2)$, for $m$ even
(resp. $(r,s)=((m-1)/2,(m+1)/2)$, for $m$ odd), is:
\be
\begin{array}{llllll}
m\ {\rm even}: && \f_0^k & 
\sim\ \phi_{2k+1,2k+1} & & k=0,1,\dots,\frac{m-2}{2},\\
               && \f_0^{(m-2)/2+k} & 
\sim\ \phi_{2k+1,2k} && k=1,2,\dots,\frac{m-2}{2},\\
&&&\\
m\ {\rm odd}:  && \f_0^k & 
\sim\ \phi_{2k+1,2k+1} && k=0,1,\dots,\frac{m-3}{2},\\
               && \f_0^{(m-1)/2+k} & 
\sim\ \phi_{2k+2,2k+1} && k=0,1,\dots,\frac{m-3}{2}.
\end{array}
\label{b-f}
\ee
The identification of the relevant fields of the other boundaries
is more subtle and will be discussed for specific
cases in the next Section.

In the rest of this Section, we show how to compute the
ground state field profile at bulk criticality and its 
classical action (free energy) for each solution of the algebraic
equation (\ref{alg-eq}).
The profile is obtained from Eq.(\ref{sol-eq}) with critical bulk potential
(\ref{LGbulkV}) ($\l=1/2$): 
\be
\f_{\rm sol}(x)\ =\ 
\mp\ \left[\left(m-2\right)\left(x+C \right)\right]^{-1/(m-2)}\ .
\label{f-prof}
\ee
The integration constant $C$  should be positive 
in order to avoid singularities at finite $x>0$.
Furthermore, we should choose the branch that monotonically 
approaches $\f=0$ for $x\to +\infty$: this relates the sign of the boundary
field value, $\f_o=\f_{\rm sol}(0)$, with that of the boundary derivative,
as follows:
\be
0 \ >\   \f_o \left.\frac{\partial\f_{\rm sol}}{\partial x}\right\vert_0 
\ = \ \f_o\ \frac{\partial V_b}{\partial\f_o}\ = \pm\ \f_o^m \ . 
\label{eq-sign}
\ee
Here we used both equations (\ref{full-bc-eq}) to determine the
sign of the derivative.

Equation (\ref{eq-sign}) fixes the free sign in 
Eq.(\ref{alg-eq}) to be $(-)$ for 
$m$ even, and $(-)$ for $\f_o>0$ (resp. $(+)$ for $\f_o<0$) for $m$ odd. 
The function $W$ is thus completely specified:
\ba
W\!\! &=&
\!\!\int^{\f_o}_0 d\f \sqrt{2V_{crit}}\ + V_b\ \ = \ \frac{\f_o^m}{m}\ +\ 
a_{m-3} \ \frac{\f_o^{m-2}}{m-2} \ +\ \cdots \ +\ a_0\ \f_o \ , 
\quad (m\ {\rm even});\nl
W\!\! &=
&\!\!\int^{\f_o}_0 d\f  \left\vert\sqrt{2V_{crit}}\right\vert + V_b\ =
\left\vert\frac{\f_o^m}{m}\right\vert +\ 
a_{m-3} \ \frac{\f_o^{m-2}}{m-2} \ +\ \cdots \ +\ a_0\ \f_o \ , 
\quad (m\ {\rm odd}).\nl
\label{w-stat}
\ea

The reduction of the boundary problem to the stationary condition
for $W$ is not surprising, because
this quantity is actually proportional to the classical action evaluated
on the soliton solution. Indeed, following the standard steps for
computing the soliton energy, we find:
\ba
S\left[\f_{\rm sol} \right] &=& T\ E\left[\f_{\rm sol} \right] \ 
+\ T\ V_b\left(\f_o \right)\nl
&=& T\int_0^\infty dx\left(\partial_x\f_{\rm sol}\right)^2\ +\ 
T\ V_b\left(\f_o \right) = \ T\ W\left(\f_o\right) \ .
\label{sol-act}
\ea
Note that the soliton energy can be written as a boundary term
of the same type as the contribution from the boundary potential. 

The occurrence of the non-analytic modulus function in  $W(\f_o)$ 
for $m$ odd, Eq.(\ref{w-stat}), might seem to spoil the reparametrization
invariance advocated in the study of its stationary points:
$\f_o=0$ is a special point on the line, and it breaks translation invariance
that is part of the reparametrization group.
As a consequence, the critical points would be characterized by
their degeneracies and relative positions, as well as by their positions
with respect to $\f_o=0$.
Actually, this is not the case: for odd $m$,
the stationary equations for $W(\f_o)$ amount to a pair of
reparametrization-invariant problems,
\be
\begin{array}{lllr}
{\bf I} &:\ 0=+\ \f_o^{m-1}\ +\ 
a_{m-3} \ \f_o^{m-3} \ +\ \cdots \ +\ a_1\ \f_o\ +\ a_0\ , && \f_o >0\ ,\\
{\bf II} &: \ 0=-\ \f_o^{m-1}\ +\ 
a_{m-3} \ \f_o^{m-3} \ +\ \cdots \ +\ a_1\ \f_o\ +\ a_0 \ , && \f_o <0\ ,
\end{array}
\quad (m\ {\rm odd}).
\label{two-eq}
\ee
to be combined together with the $\f_0$ restriction. 
Each equation allows the identification of the solutions with the
boundaries according to the reparametrization-invariant rules discussed before;
moreover, the second equation is equal to
the first one at the reflected point $(-a_1,\dots,-a_{m-2})$
of the parameter space.
Thus, we can study a single equation at each point $(a_1,\dots,a_{m-2})$,
say Eq. ${\bf I}$, first identify the solutions to the boundaries, then
keep the positive solutions $\f_o>0$, 
and get the negative ones from the reflected point.
As the examples of the next Section will clarify,
the reflection of points by the origin in the parameter space 
$(a_1,\dots, a_{m-2})$ will make it different from the moduli space
of $A_{m-1}$ singularities, but nonetheless
the singularities remain well identified: ambiguities can only arise
for the solution $\f_o=0$, but they  can be resolved 
by relying on continuity from $\f_o=0^+$ and $\f_o=0^-$.


\section{Boundary renormalization group flows}

We have seen that the most unstable boundary
is represented by the stationary point 
$\partial W/\partial\f_o=\f_o^{m-1}=0$
in Eq.(\ref{alg-eq}), i.e. by vanishing boundary potential. 
The RG flows from this point are described by switching on the
terms in $V_b$: it results into one or several 
stationary points of lower degeneracy, 
whose nature depends on the parameters $(a_1,\dots,a_{m-2})$.
One (or some) of these new stationary points will be identified with
the boundary at the infrared point of the RG flow.
Further detuning of the latter can continue the flow
until the stable boundaries (non-degenerate stationary points of $W$)
are reached.
Therefore, the boundary RG flows are certain motions in 
the $\mathbb{R}^{m-2}$ moduli spaces of deformations
of Arnold's singularities.

\subsection{Selection rule}

Before entering into the detailed analysis of some examples of these spaces,
we can derive a selection rule for the flows.
In the LG picture, a degenerate singular point is made
by collapsing neighbor points of lower singularity;
these break apart by the opposite move of detuning,
corresponding to the RG flow.
We can represent the detuning in terms of the breaking of
the Dynkin diagram associated to the singular point.
First describe the diagrams by the coordinates of their ending
points, $(n_1,n_2)$, $1 \le n_1 \le n_2 \le m-1$; 
let $(n_1,n_2)$ be the diagram of the UV boundary, and let us
consider the RG flow ending into a superposition of IR boundaries,
whose diagrams have coordinates
$\left(m_1^{(\a)},m_2^{(\a)}\right)$ for some $\a$ values.
The breaking of the UV Dynkin diagram into proper sub-diagrams
imply the following bounds on the coordinates of the IR diagrams:
\be
(n_1,n_2) \ \longrightarrow \ \bigoplus_\a\ 
\left(m_1^{(\a)},m_2^{(\a)}\right)\ , \qquad 
n_1\le m_1^{(\a)} \le m_2^{(\a)} \le n_2\ .
\label{rg-rule}
\ee
This selection rule  implies that the boundary RG flows
form nested patterns.

We can use this rule to identify the conformal boundaries one-to-one with
the Dynkin diagrams, i.e. relate the indexes $(n_1,n_2)$ to
the Kac labels $(r,s)$ of the boundaries.
The $(m-1)$ stable boundaries $(r,1)$, $r=1,\dots,m-1$, are naturally ordered 
in $r\sim height$ (also confirmed by the $\Z_2$ action $r\to m-r$), 
as much as the non-degenerate stationary points on the real line:
thus, $n_1=n_2=r$  for $s=1$.
The unstable boundaries can be localized by using their RG
flows to superpositions of stable boundaries under the
$\phi_{(1,3)}$ perturbation\footnote{
The following formulae do not respect the reflection symmetry of the
Kac table, and thus strictly hold for $r,s \ll m$.}, that has been
discussed in the Refs.\cite{rrs,watts,kevin}:
\ba
(r,s) \ +\phi_{(1,3)}&\longrightarrow &
\bigoplus_{\ell=1}^{min(r,s)} \left(r+s+1-2\ell,1 \right)\ ,\nl
(r,s) \ - \phi_{(1,3)}&\longrightarrow &
\bigoplus_{\ell=1}^{min(r,s-1)} \left(r+s-2\ell,1 \right)\ .
\label{phi13-flow}
\ea
In these flows the UV Dynkin diagram gets broken into its 
smaller pieces, the points, that all occur once.
Using the flow selection rule, the position of the UV diagram is tied
to that of its points, whose Kac indexes where already identified.

The result of the matching is, for boundaries with {\it even}
number of relevant fields ($1 \le n_1 \le n_2 \le m-1$):
\be
s=\frac{n_2-n_1}{2}+1\ ,\qquad r=\frac{n_2+n_1}{2}\ ,\qquad
(s \le r\le m-s)\ ;
\label{map-even}
\ee
and for an {\it odd} number of relevant fields,
\be
s=\frac{n_2-n_1+1}{2}\ ,\qquad r=\frac{n_2+n_1+1}{2}\ ,\qquad
(r < s < m+1-r)\ .
\label{map-odd}
\ee
In Ref.\cite{kevin}, a similar description of the conformal boundaries
in terms of Dynkin diagrams has been proposed, 
that is based on the relation between
CFT and lattice integrable models \cite{pearce}.
The $m$-th Virasoro model was actually related to the $A_m$ Dynkin
diagram, rather than the $A_{m-1}$ one considered here:
however, the two diagrams are related by the lattice ``duality'',
mapping the points of the $A_{m-1}$ diagram to the bonds of the $A_m$ one.
After duality, one finds that the two proposed 
identifications diagrams-boundaries agree completely.
In Ref. \cite{kevin}, the selection rule of breaking the
diagrams was already observed in some examples of perturbative RG
flows  ($m\to\infty$): our results generalize these findings.

Moreover, the opposite move of joining diagrams were
also observed in Ref.\cite{kevin}
for flows starting from superpositions of Cardy states,
for example (in diagram coordinates):
\be
(n-2,n-1)\ \oplus\ (n,n+1) \longrightarrow (n-2,n+1) \ , \qquad
(n \ll m\to\infty). 
\label{graham-join}
\ee
Further flows from that IR point anyhow corresponded to splittings, e.g.
\be
(n-2,n+1) \longrightarrow
(n-2,n-1)\ \oplus\ (n+1,n+1) \longrightarrow (n-1,n-1)\ \oplus \ (n+1,n+1)\ .
\label{graham-split}
\ee
In the flows (\ref{graham-join}) the two UV diagrams are joined by adding
a bond between neighbor points: this amounts to a deformation of $W$ 
bringing two stationary points together; such fine-tunings are forbidden 
for critical phenomena, but allowed for first-order phase transitions.
Therefore, these flows are likely to be weak first-order transitions,
not distinguishable from second order in the perturbative regime,
that are driven by the second identity field present
in the UV superposition of boundaries.

Let us finally mention another labelling
of boundary states by {\it pairs} of Dynkin diagrams.
In  Ref.\cite{zuber}, the boundary states of the Virasoro minimal models
have been completely classified: for any model characterized by
the pair of diagrams $(A_{h-1},G_n)$, with $G=A,D,E$, the boundary states are
labelled by the indexes $(r,a)$, denoting the nodes of the respective
diagrams in the pair\footnote{
For the diagrams $G=A,D_{\rm odd},E_6$ presenting 
a $\Z_2$ reflection symmetry, $a\to\g(a)$, the boundaries
should be identified in pairs, $(r,a)\sim(h-r,\g(a))$ .},
$r=1,\dots,h-1$ and $a=1,\dots,n$.
In the $(A,A)$ series considered so far, the boundaries are associated
to the nodes of one $A$ diagram and have an extra number attached to it
(for the other diagram), while we used both nodes and bonds of a single
diagram.


\subsection{Renormalization group
 spaces versus moduli spaces of $A_{m-1}$ singularities}

\subsubsection{Tricritical Ising model and $A_3$ space}

\begin{figure}[ht]
\begin{center}
\includegraphics[width=9cm]{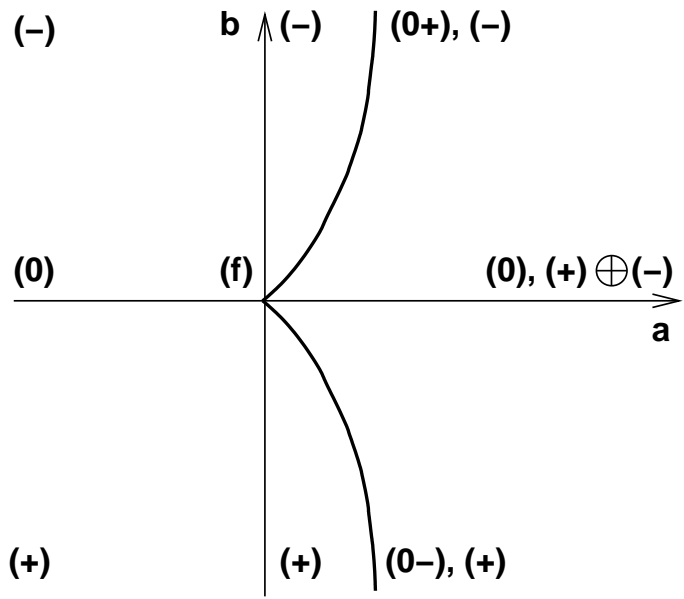}
\end{center}
\caption{LG description of the boundary RG flows in the Tricritical Ising
model.} 
\label{fig-cusp}
\end{figure}

The algebraic equation for the LG boundary condition (\ref{alg-eq}) 
corresponds to the deformations of the $A_3$ singularity ($m=4$),
\be
0=\ \frac{\de W}{\de x}\ =\ x^3 \ -\ a\ x \ +\ b\ .
\label{tric-eq}
\ee
This gives rise to the two-dimensional parameter space $(a,b)$
shown in Figure \ref{fig-cusp}.
The parameter $a$ is parity even, while $b$ is odd; the two-fold
degenerate stationary point sits at the origin and the once
degenerate points live on the wings of the cusp,
\be
\left(\frac{b}{2} \right)^2 = \left(\frac{a}{3} \right)^3\ ,
\qquad \qquad x_o^I= -2 \left(\frac{a}{3} \right)^{1/2}\ , \quad
x_o^{II}= \left(\frac{a}{3} \right)^{1/2}\ ,
\label{cusp-eq}
\ee
where $x^I_o$ and $x^{II}_o$ denote the positions of the
non-degenerate and degenerate stationary points, respectively.
Three non-degenerate solutions exists to the right of the wings and
one to the left.
It may be useful to schematically draw the shape of the free
energy $S(\f_o)=T\ W(\f_0)$ in each region of the $(a,b)$ plane.
The stationary points of $W$ can be identified with the
Tricritical Ising conformal boundaries by using the rules
in Figure \ref{fig-dynkin} and paying attention to the continuity 
of solutions w.r.t. parameter changes (see Figure \ref{fig-cusp}).

Let us start by discussing the deformations of the most unstable
boundary $(f)$ and compare them
with the known RG flows summarized in Figure \ref{fig-tricri}.
The parity-odd flows,
\be
(f)\ \pm\ \phi_{1,2}\ \longrightarrow\ (\pm) \ ,
\label{odd-rg}
\ee
match the deformation to a single non-degenerate critical point made
by the $b$ term, $\de W/\de x = x^3 + b=0$.

The parity even flows, 
\be
(f)\ \ -\ \phi_{1,3}\ \longrightarrow\ (0)\ ,\qquad
(f)\ \ +\ \phi_{1,3}\ \longrightarrow\ (+) \oplus (-)\ .
\label{even-rg}
\ee
correspond to the even $a$ deformation, $x^3 -a x=0$.
Actually, for $a>0$ all three stable stationary
points appear and we should understand which ones are chosen for the 
IR fixed point.

Here, we should remark an important difference with respect to
the usual mean field theories \cite{binder}.
In the presence of spontaneous symmetry breaking, the potential
barriers separating the different stationary points 
get renormalized to infinity by volume effects,
such that any minimum of the potential corresponds to a stable ground state.
In the present case of absence of spontaneous symmetry breaking,
the tunnelling barriers remain finite and thus
none of the minima (and maxima) of the effective action 
are stable at the semi-classical level.
One stationary point can only be singled out
by tuning some of the parameters of $W$ to infinity: therefore,
the IR fixed points of the RG flow should be
found at infinity in the $(a,b)$ parameter space, taking limits along 
certain curves while selecting specific stationary points.
Such limits also send $\f_o\to\infty$ (with specific speed)
in agreement with conformal invariance.

The identification of the conformal boundaries (fixed points) with the
stationary points of $W$ reported in Figure \ref{fig-cusp} indeed
holds for asymptotic values of the $(a,b)$ parameters.
It is thus consistent to match the unique stationary point on the left
of the wings to three different boundaries, namely $(+),(0),(-)$, 
in the respective asymptotic regions:  
$(a,b)=(-\infty,-\infty), (-\infty, 0), (-\infty +\infty)$.

In conclusion, the second RG flow in Eq.(\ref{even-rg}) is obtained by letting 
$a\to +\infty$ and sitting on the degenerate pair of minima $(+),(-)$.
The superposition of Cardy boundaries survives the IR limit
because the $\Z_2$ symmetry is preserved at all stages.

The plot of $W$ along the wing $b>0$ shows a degenerate stationary point 
corresponding to $(0+)$ and the stable minimum $(-)$.
The mixed flow,
\be
(f) \ +\l\ \phi_{1,2} \ +\m\ \phi_{1,3}\ \longrightarrow\ (0+)\ ,
\label{mix-rg}
\ee
is thus reproduced by going to infinity along the wing while sitting
on $x=x_o^{II}(a,b)$ of $W$.
These examples show that the LG description of the flows involves
both a choice of deformation parameters, such as $(a,b)$,
and a choice of ground state; thus,
the deformation parameters can be related to the coupling
constants of the perturbed CFT, such as $\l,\m$, but each flow
requires specific identifications.

We now perform the limit $a,b\to\infty$ on the wing $b>0$ for describing the
further flows:
\be
(0+) \ +\ \g\ \phi_{1,3}\ \longrightarrow\ (0)\ ,\qquad\qquad
(0+) \ -\ \g\ \phi_{1,3}\ \longrightarrow\ (+)\ .
\label{sub-rg}
\ee
Using parameters that blow up the region around the wing,
\be
b= 2\left(\frac{a}{3} \right)^{3/2} + \wt{b}\ ,\qquad
x = \left(\frac{a}{3} \right)^{1/2} + \wt{x}\ , \qquad\qquad
a \gg \wt{x},\wt{b}\ ,
\label{blow-up}
\ee
we find:
\be
 W(x) \ =\ {\rm const.}\ +\ \frac{\wt{x}^4}{4}\ +
\left(\frac{a}{3}\right)^{1/2}\ \wt{x}^3\ +\ \wt{b}\ \wt{x}\ .
\label{blow-res}
\ee
In the limit $a\to\infty$, we can neglect the highest power, reducing the
problem to the deformation of the $A_2$ singularity $\wt{x}^3=0$
already discussed in the Ising case.
However, we should pay attention to the convexity of the scaled
free energy, $\wt{W}\sim \wt{x}^3 +\wt{b} \wt{x}$, at large $\wt{x}$:
we can take the limit $\wt{x}\to+\infty$, but not to $-\infty$.
We can cure this problem by further translating $\wt{x}$ for $\wt{b}<0$, 
such that one of the two stable stationary points always
sits at $\wt{x}=0$ (the $(0)$ boundary) and the other one $(+)$ is at
$\wt{x}_o>0$.


\subsubsection{Tetracritical model and the ``swallow tail'' singularity}

\begin{figure}[ht]
\begin{center}
\includegraphics[width=16cm]{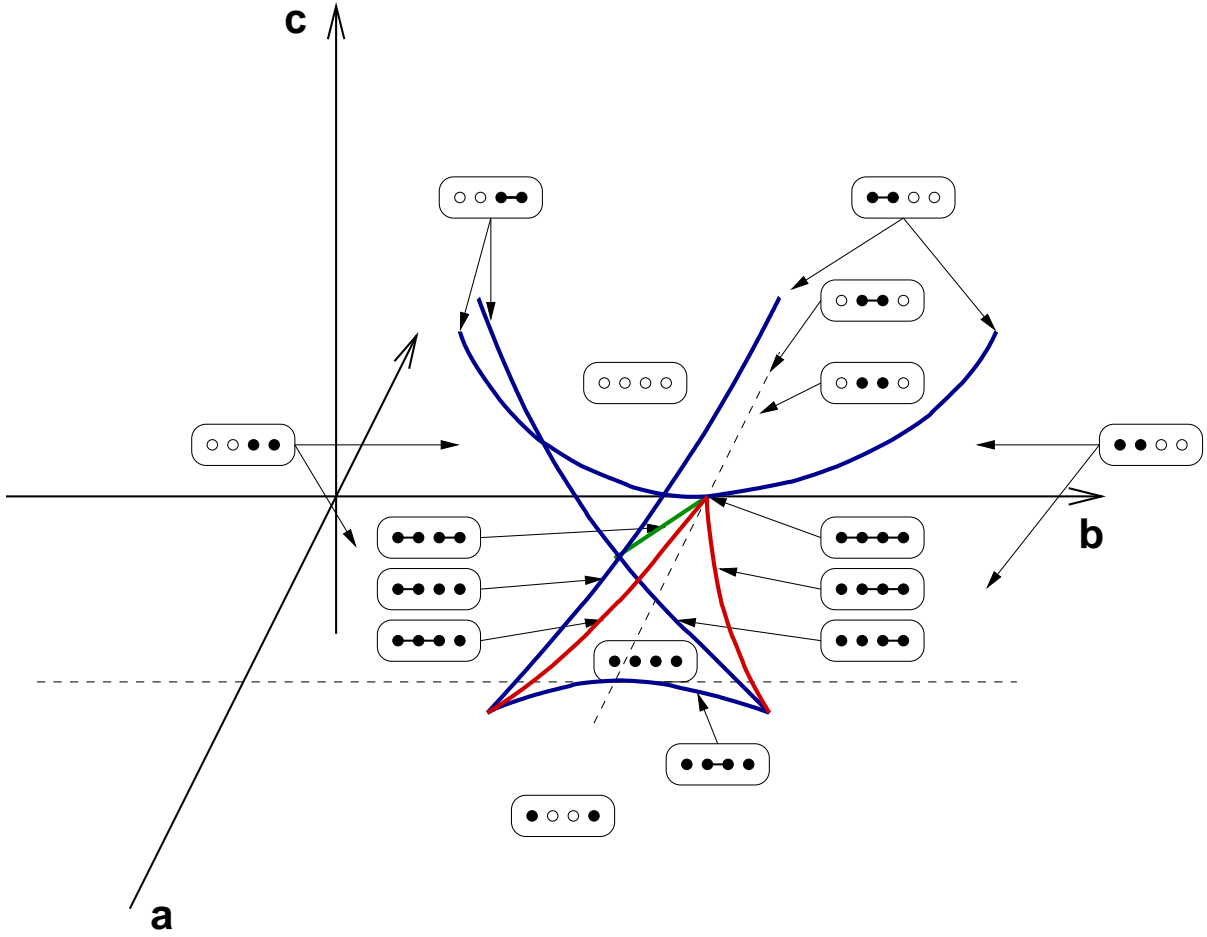}
\end{center}
\caption{$A_{4}$ parameter space: the Dynkin sub-diagrams of the
singularities are drawn in each region; open dots indicate the
missing (complex) roots and are reported for locating 
the sub-diagrams w.r.t. the $A_4$ diagram.}
\label{fig-tail}
\end{figure}

The stationary equations (\ref{two-eq}) for the $m=5$ LG theory are:
\be
\begin{array}{lllr}
{\bf I} &:\   0=+\f_o^4 +a_2\ \f_o^2 + a_1\ \f_o +a_0 \ , && \f_o >0\ ,\\
{\bf II} &: \ 0=-\f_o^4 +a_2\ \f_o^2 + a_1\ \f_o +a_0 \ , && \f_o <0\ .
\end{array}
\label{tail-eq}
\ee
We shall first discuss the parameter space of the $A_4$ singularity,
\be
0=x^4 + a\ x^2 +b\ x +c\ ,
\label{tail-sing}
\ee
and later implement the extra conditions in (\ref{tail-eq}).  
The $A_4$ parameter space is drawn in Figure \ref{fig-tail},
together with the Dynkin diagrams associated to the singularities in
all the regions.
For $a>0$ a surface roughly orthogonal to the $c$ axis divides the
regions with two and no singular points, respectively.
The figure is symmetric with respect to the plane $b=0$.
For $a<0$ the surface folds in a way similar to the tail of the swallow
and creates a three-sided wedge with the tip at the origin of the axis;
the region inside the wedge contains four non-degenerate singularities.
The two lower edges of the wedge, with equations,
\be
b^2=\left(\frac{-2a}{3} \right)^{3/2} \ , \qquad\qquad
c=- \frac{a^2}{12}\ , \qquad\qquad (a<0)\ ,
\label{tail-wing}
\ee
extend in the region $c<0, a<0$, and 
locate the two-fold degenerate singularities; the three surfaces
of the wedge contain a once degenerate singularity and two non-degenerate ones.
On the curve at the pinching of the surface, there are two once
degenerate singularities, one of which disappears in the surfaces above 
the pinching.

\begin{figure}[ht]
\begin{center}
\input{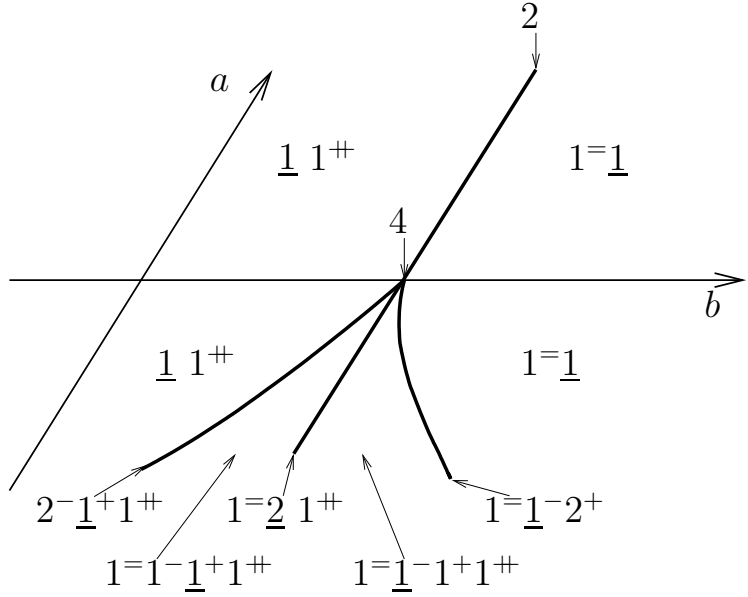}
\end{center}
\caption{$c=0$ plane of the $A_{4}$ parameter space, with identification of
the singularities using continuity from the neighbor regions
$c=0^+$ and $c=0^-$.}
\label{fig-plane}
\end{figure}

\begin{figure}[ht]
\begin{center}
\input{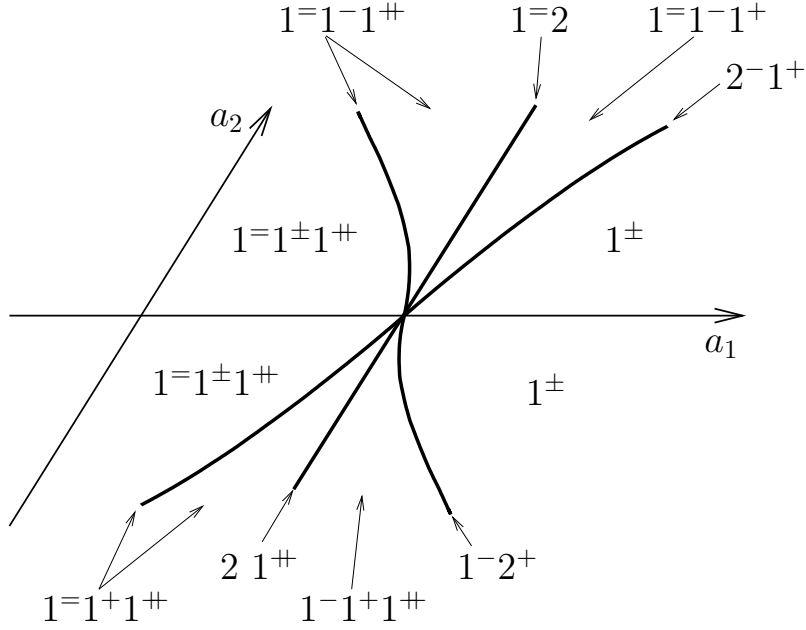}
\end{center}
\caption{RG flow space of the Tetracritical Ising model for $a_0=0$:
it has been obtained from Figure \ref{fig-plane} by reflecting 
the negative solutions as explained in the text;
the boundary $(1^+)\oplus(1^-)$ is indicated by $(1^\pm)$.}
\label{fig-phys}
\end{figure}

The nature of the stationary points can be unambiguously identified 
in all the regions of the parameter space; besides their Dynkin diagram, 
let us characterize the points by the following names: 
\be
(4)\ ,(3^-)\ ,(3^+)\ ,(2^-)\ ,(2)\ ,(2^+)\ ,(1^=)\ ,(1^-)\ ,
(1^+)\ ,(1^{\+})\ ,
\label{tail-num}
\ee
where the number denotes the degree of degeneracy plus one, 
and the plus and minuses the ordering of the points on the real line.

Let us proceed to mapping the solutions
of equation (\ref{tail-sing}) with those of (\ref{tail-eq}). 
First we remark that
the reflection condition for $m$ odd (\ref{two-eq}) implies that the 
physical parameters in $V_b$ have opposite $\Z_2$ parities than those of the
$A_4$ singularity: $a_1$ is parity invariant, while
$a_2$ and $a_0$ are partially and completely parity breaking, respectively;
on the other hand, looking to the Eq.(\ref{tail-sing}) we see that
the $a$ and $c$ deformations are even and the $b$ one is odd.

The flow obtained by switching on the
parameters $a_2,a_1,a_0$ in $V_b$ is found by solving the system
(\ref{tail-eq}) as follows: we consider the $A_4$
singularities (\ref{tail-sing}) at the same point, $(a_2,a_1,a_0)=(a,b,c)$,
select the positive solutions, $\f_o=x_o>0$, and discard the negative ones; 
then, we consider the reflected point, $(a_2,a_1,a_0)=(-a,-b,-c)$,
and take the negative solutions of (\ref{tail-sing}).
For example, consider the $\phi_{1,3}$ flow 
discussed earlier, Eq. (\ref{phi13-flow})  \cite{rrs}:
\be
(4)\ +\ \phi_{1,3} \ \longrightarrow\ (1^-)\ \oplus\ (1^{\+})\ ,\qquad\quad
(4)\ -\ \phi_{1,3} \ \longrightarrow\ (1^=)\ \oplus\ (1^+)\ .
\label{pert-rg}
\ee
It is reproduced by moving along the $a$-axis inside the wedge, 
$a_2=a<0$, $a_1=b=0$, $a_0=c=0^+$: the positive singularities 
$(1^+),(1^\+)$ are kept and the negative ones,  $(1^=),(1^-)$, 
are neglected; instead, from the reflected point 
$a=-a_2>0,b=-a_1=0,c=-a_0=0^-$,
$(1^-)$ is taken and $(1^+)$ discarded.
The superposition of IR boundaries in the first of Eqs.
(\ref{pert-rg}) is then reproduced: $(1^{\+})$ and $(1^-)$ 
correspond to two minima of $W$ that can be tuned 
to the same height.

The solutions involving $\f_o=0$ lay on the $c=0$ plane and need a 
detailed discussion. The specialization of the $A_4$ parameter space 
(Figure \ref{fig-tail}) to the $c=0$ plane is drawn in 
Figure \ref{fig-plane}.
The stationary points of $W$ are identified with the conformal boundaries
using continuity from the regions $c=0^+$ and $c=0^-$:
the underlined symbols correspond to the solution $\f_o=0$.
One ambiguity occurs near the $b$ axis, where the vanishing
solution indicated by $(\und{1})$ is the limit of $(1^+)$ from 
$c=0^+$ and of $(1^-)$ from $c=0^-$: it can be interpreted as 
one of the two cases or as their parity invariant combination
$(1^+)\oplus (1^-)$.

This ambiguity is resolved after reflection of the negative solutions,
as required by Eq. (\ref{tail-eq}), leading to Figure \ref{fig-phys}.
On the $a_1$ axis, corresponding to 
$W=\left\vert \f_o^5/5 \right\vert + a_1 \f_o^2/2$,
one should take the parity invariant combination, 
$(\und{1})\sim (1^+)\oplus (1^-)$, and in the neighbor regions as well.
Inside the wings, the identification is different,
$(\und{1})\sim (1^+)$ or $(\und{1})\sim (1^-)$,  
by comparing with the reflected solutions.
In conclusion, the analysis of the $A_4$ theory confirms that the
potential ambiguities in the interpretation of the $\f_o=0$ solution
can be resolved by using continuity arguments.


\subsection{Remarks on the (A,D) and (A,E) series}

The Landau-Ginzburg description of the bulk Virasoro models
in the (A,D) and (A,E) series has been discussed in the 
Refs.\cite{zamlg}\cite{kekeli}.
The best known example is the three-state Potts model $(A_4, D_4)$ for $m=5$,
that has been described by a two-component scalar theory with 
$\Z_3$-symmetric potential:
\be
V_{\rm crit} \left(\f_1,\f_2 \right)= 
\l\ \R\left(\f_1 \ +\ i\ \f_2\right)^3 \ .
\label{v-pot}
\ee
The five relevant fields of the Potts model,
$\s,\ov{\s},\varepsilon, \psi, \ov{\psi}$,
 have been identified as follows:
\be
\s\ = \ \f_1 \ +\ i\ \f_2\ , \qquad
\varepsilon\ =\ \f_1^2\ +\ \f_2^2\ , \qquad
\psi =  \left(\f_1 + i\f_2\right)\left(\f_1^2 + \f_2^2\right)\ ,
\label{f-pot}
\ee
where the first and third fields are complex and the second one is real.
On the other hand, there are three deformations
modulo reparametrizations of the Arnold $D_4$ singularity, with
real parameters $a,b,c$:
\be
W\ = \ x^3\ -\ 3\ x\ y^2 \ +\ a\ x\ +\ b\ y\ 
+\ c\left(x^2 \ +\ y^2\right) \ .
\label{d-sing}
\ee
Actually, after the identification $(\f_1 \sim x, \f_2\sim y)$,
the fermion fields $\psi$, $\ov{\psi}$ 
corresponds to deformations of $W$ under 
reparametrizations of the $(x,y)$ plane, of the form 
$\de W/\de x \ f + \de W/\de y\ g $.
Note also that the scaling dimension of $\psi$ is two, 
i.e. it is marginal at the classical level, but it becomes relevant
at the quantum level due the extra anomalous dimension
acquired by composite fields.

In conclusions, the quantum theory of the Potts model 
possesses more relevant fields that those occurring in the 
semiclassical LG theory, thus signaling a violation of
reparametrization invariance.
The LG classical theory is only exact in the supersymmetric case \cite{vw},
where also composite fields do not acquire extra anomalous dimensions.
These renormalization effects are also present in 
the other non-supersymmetric (A,D) and (A,E) Virasoro models that 
presents roughly twice relevant fields than those found from the
reparametrization-invariant deformations of the LG potential\footnote{
The dimensions of the ring of deformations can be computed in all cases
by using a formula given in Ref.\cite{lvw}.}.
It turns out that the bulk LG description is nevertheless useful
for describing the CFT results on field symmetries and fusion rules
\cite{kekeli}.

On the contrary, the LG description of boundary states in 
this paper is tied to the hypothesis of reparametrization invariance
of deformations of Arnold's singularities and is affected by 
the mismatch in the set of relevant fields.
The Potts model boundaries are described in Appendix A:
the deformations of the $D_4$ singularity do reproduce
all boundary states of the model,
but their degree of instability is lower, lacking the pair of
the relevant deformations corresponding to $\psi,\ov{\psi}$.
The resulting reduced, three-dimensional 
space of boundary RG flows is nevertheless consistent with the known results
from CFT and integrable models \cite{potts}.

\setcounter{footnote}{0}

\section{N=2 supersymmetric Landau-Ginzburg theory and minimal models}

The minimal models of N=2 superconformal symmetry have received a lot
of attention over the years due to their relevance for String Theory model
building \cite{gepner} (for a review see \cite{greene}).
In the last few years, the allowed supersymmetric boundary conditions have been
found \cite{rs}
and several papers have investigated their geometric interpretation
\cite{quintic} \cite{mms}.
As shown in the paper by Hori, Vafa and Iqbal \cite{vafa}, 
the Landau-Ginzburg theory can provide an interesting 
geometrical description of the boundary conditions, building on the
rich mathematical properties of the bulk theory already established in 
Ref. \cite{cecotti}.
After introducing these results, we shall show that our
LG description of the N=0 boundary conditions 
can be extended to the supersymmetric case, where it is found
to be equivalent to the Hori et al. analysis.
We shall then conclude with some observations on the supersymmetric
boundary RG flows that are quite different from the N=0 case.

\subsection{Introduction: BPS solitons and A-type boundary conditions}

The N=2 minimal conformal field theories can be realized by the coset
$\widehat{SU(2)}_k/\widehat{U(1)}_4 \otimes \widehat{U(1)}_{2k+4}$,
with central charge $c=3k/(k+2)$, $k=1,2,\dots$.
Their bulk sectors are labelled by the triples $(\ell,m,s)$,
with $m$ modulo $2(k+2)$ and $s$ modulo $4$,
subjected to one identification and one condition,
\ba
&&\ell = 0,1,\dots,k\ ,\qquad m=-k-1,\dots,k+2\ , \qquad s=-1,0,1,2\ ;\nl
&&(\ell,m,s) \sim (k-\ell,m+k+2,s+2)\ ,\nl
&&\ell+m+s=0\ \ {\rm mod}\ 2\ ,
\label{susy-no}
\ea
such that there are $(k+2)(k+1)$ states both in the Neveu-Schwarz (NS)
($s=0,2$) and in the Ramond (R) ($s=1,-1$) sectors.
The dimensions and $U(1)$ charges of the superconformal fields are:
\be
h^\ell_{m,s}\ =\ \frac{\ell(\ell+2)-m^2}{4(k+2)}\ +\ 
\frac{s^2}{8}\ +\ \Z\ ,\qquad
q^\ell_{m,s}\ =\ \frac{m}{k+2}\ -\ \frac{s}{2} \ +\ 2\ \Z\ .
\label{fields}
\ee
Each minimal model is characterized by a single A-D-E Dynkin diagram:
we shall only deal with the simplest models of the A series 
that contains all the bulk sectors once.

The N=2 supersymmetry is generated by the four supercurrents
$G_\pm, \ \ov{G}_\pm$ whose zero modes are the supercharges
$Q_\pm, \ov{Q}_\pm$.
The boundary conditions can respect
N=2 supersymmetry in two different ways, called (A) and (B) types,
corresponding to the following linear combinations:
\ba
(A):&& Q=\ov{Q}_+ +\eta\ Q_-\ , \qquad Q^\dagger=\ov{Q}_- +\eta\ Q_+\ ,
\qquad \eta=\pm 1\ ,\nl
(B):&& Q=\ov{Q}_+ +\eta\ \ov{Q}_-\ , \qquad Q^\dagger=Q_+ +\eta\ Q_-\ ;
\label{ab-bc}
\ea
these amount to real and holomorphic identifications of the
infinitesimal supersymmetric parameters,
(A): $\eps_-=\eta\ov{\eps}_+$ and (B): $\eps_-=-\eta \eps_+$, respectively.
The signs $\eta=- 1,1$ pertain to the NS and R sectors, respectively. 

The (A)-type boundary states can be interpreted as D1 branes and
the (B)-type ones can be either D2 or D0 branes \cite{rs}\cite{mms}.
The (A)-type boundary states were described by Hori et al. \cite{vafa}
 in terms of the BPS solitons
of the LG theory\footnote{
The D2 branes have also been described in the LG theory by adding
degrees of freedom at the boundary \cite{lerche}.}.
In the conformal theory, the (A) boundaries are (generalized) 
Cardy states and are labelled
by the same quantum numbers $(\ell,m,s)$ of the bulk sectors. 
The basic generators of the
$\Z_{k+2}$ and $\Z_2$ symmetries of the theory act on these states 
by shifting $m\to m+2$ and $s\to s+2$, respectively.
Owing to the symmetries of their labels, the boundary states
of the Ramond sector, $s=\pm 1$, can be associated to the following 
diagrams: draw a regular polygon with $k+2$ vertices, 
$z_n=\exp(i2\pi n/(k+2))$, inside the unit circle; the oriented sides and
all the chords of the polygon, between pairs of vertices $(n_1,n_2)$ mod $k+2$,
represent the $(k+2)(k+1)$ boundaries according to the following 
identification\footnote{
A drawing with $2k+4$ points on the circle can also be used to
represent both NS and R boundaries \cite{mms}.},
\ba
\ell +1 &=& \left\vert n_2-n_1 \right\vert \ ,\qquad\qquad
 0\le n_1,n_2 \le k+1\ , \nl
m &=& n_1+n_2\ , \nl
s &=& {\rm sign}(n_2-n_1)\ ,\nl
g_{\ell,m,s}&=& g_\ell\propto \sin\frac{\pi(\ell+1)}{k+2}\ .
\label{polyg}
\ea
Note that the boundary entropy $g_{\ell,m,s}$ 
is proportional to the length of the chords.
We remark that the interplay with the N=0 models is through $m\sim k+2$:
indeed, the $(n_1,n_2)$ labelling of Dynkin (sub)-diagrams
proposed for the N=0 boundary states in Section 3 (Figure \ref{fig-dynkin})
becomes equal to that of the  N=2 boundary states (\ref{polyg})
upon replacing the $A_{k+1}$ Dynkin diagram
with that of the corresponding affine Lie algebra $\wh{A}^{(1)}_{k+1}$,
which contains one extra point and two bonds to close the chain into
a polygon.

The bulk N=2 Landau Ginzburg theory is described by the action 
\cite{vafa}:
\be
S=\int d^2x\left[ d^2\th\ d^2\ov{\th}\ K\left(\F,\ov{\F}\right)\ +\ 
\int d^2\th\ {\cal W}\left(\F\right)\ +\ 
\int d^2\ov{\th}\ \ov{\cal W}\left(\ov{\F}\right)\right]\ ,
\label{susy-act}
\ee
where ${\cal W}$ is the holomorphic superpotential and
the K\"ahler potential is quadratic in the chiral field $\F$ at the UV point,
$K\left(\F,\ov{\F}\right)= \F\ov{\F}$.
The theory with superpotential given by the (complex) A-D-E 
polynomial is known to flow to the corresponding A-D-E minimal model
\cite{vw}: for the $k$-th model in the A series, 
one should consider a single chiral field and the polynomial
of the $A_{k+1}$ singularity:
\be
{\cal W}_{\rm crit}=\l \ \F^{k+2}\ .
\label{susy-a}
\ee
As originally discussed in Ref.\cite{lvw},
the powers $\F^\ell$, $\ell=0,1,\dots,k,$ generate the ring of deformations
of the singularity and correspond to the chiral fields of the
N=2 algebra, $(\ell,m,s)=(\ell,\ell,0)$, the special fields
with scaling dimensions proportional to the charge that obey
a non-singular operator-product algebra; their classical LG dimensions
$h=\ell/(k+2)$ do not get renormalized.

The bulk LG theory with non-critical ${\cal W}$ possesses non-trivial 
solutions that are BPS solitons preserving half of the supersymmetry
\cite{cecotti}:
upon expanding the superfield $\F=\phi+\th\psi+\cdots$,
the energy of a soliton extending between two stationary points
of the potential, ${\cal W}(a)$ and ${\cal W}(b)$, can be written,
\ba
E &=& \frac{1}{2}\int dx \left( \de_x\phi\ \de_x\ov{\phi} \ +\ 
\frac{\de {\cal W}}{\de\phi}\frac{\de\ov{\cal W}}{\de\ov{\phi}} \right)= \nl
&=&\frac{1}{2}\int dx 
\left(\de_x\phi -\w\frac{\de\ov{\cal W}}{\de\ov{\phi}} \right)
\left(\de_x\ov{\phi} -\ov{\w}\frac{\de{\cal W}}{\de{\phi}} \right)\ +\ 
\R\left[\ov{\w}\ {\cal W}\left(\phi \right)\right]_a^b \ ,
\label{bps-en}
\ea
with $\w$ a constant phase.
The vanishing of the first term in the energy gives the soliton equation,
that possesses the first integral:
\be
\I \left[\ov{\w}\ {\cal W}\left(\phi \right)\right] = {\rm const.}\ .
\label{bps-const}
\ee
Therefore, the soliton are straight lines in the complex ${\cal W}$  plane
connecting the two stationary points ${\cal W}(a)$ and  ${\cal W}(b)$,
with slope ${\rm Arg}(\w)={\rm Arg}\left({\cal W}(b)- {\cal W}(a)\right)$;
along the line, the energy 
$\R\left[\ov{\w}\ {\cal W}\left(\phi \right)\right]$ is monotonically
increasing (or decreasing).
The type of supersymmetry preserved by the soliton can be found
by expressing the supercharges in terms of the LG fields \cite{vafa}; 
the result is that the static solitons with $\w=\pm i$ yield the 
invariant states:
\be
\de_x\phi =\pm\ i\ \frac{\de\ov{\cal W}}{\de\ov{\phi}} \quad \leftrightarrow
\quad Q\vert BPS\rangle=\ Q^\dagger\vert BPS\rangle=0\ .
\label{susy-vac}
\ee
The supersymmetric charges $Q$ and $Q^\dagger$ are actually 
the same combinations (\ref{ab-bc}) earlier
considered for the (A)-type boundary conditions\footnote{The most general 
(A) and (B) boundary conditions can also contain a free
phase but this should be conventionally fixed once for all
in defining the boundary theory.}.

The results of the Refs. \cite{vafa}\cite{maxim} have shown that this
relation between solitons and (A)-type boundary conditions is indeed general.
For the theory defined on the half space $(t,x>0)$, the 
state conditions (\ref{susy-vac}) can be read as conditions for the
boundary states at $x=0$ in the formulation in which space and time 
are interchanged, i.e. evolution takes place in $x$.
By rotating back to the ordinary setting, one finds:
\be
(A)-{\rm type\ b.\ c. } : \qquad \qquad
\de_t\phi =\pm\ \frac{\de\ov{\cal W}}{\de\ov{\phi}} 
\quad \leftrightarrow\quad \I {\cal W}(\phi) = {\rm const.}\ .
\label{susy-curve}
\ee
Therefore, the (A)-type D1 brane is a curve $\g$ in the complex
$\phi$ plane which obeys this soliton
equation and is characterized by a constant value of
the imaginary part of the superpotential\footnote{
For theories with more than one chiral field, the general result is that
the boundary is a (middle-dimensional) Lagrangian sub-manifold of the 
target space for which $\I{\cal W}$=const. \cite{vafa}.}.
The careful analysis of the supersymmetric variation of the LG action
(\ref{susy-act}) on half space has been carried out in Ref.\cite{maxim}, 
including all the supersymmetric consistency conditions; 
the result is the boundary term,
\be
\d_{\rm Susy(A)}S\propto \int_\g dt\ \de_t\left( \I{\cal W}(\phi)\right)\ ,
\label{susy-max}
\ee
that indeed vanishes on the same type of curves.

\begin{figure}[ht]
\begin{center}
\input{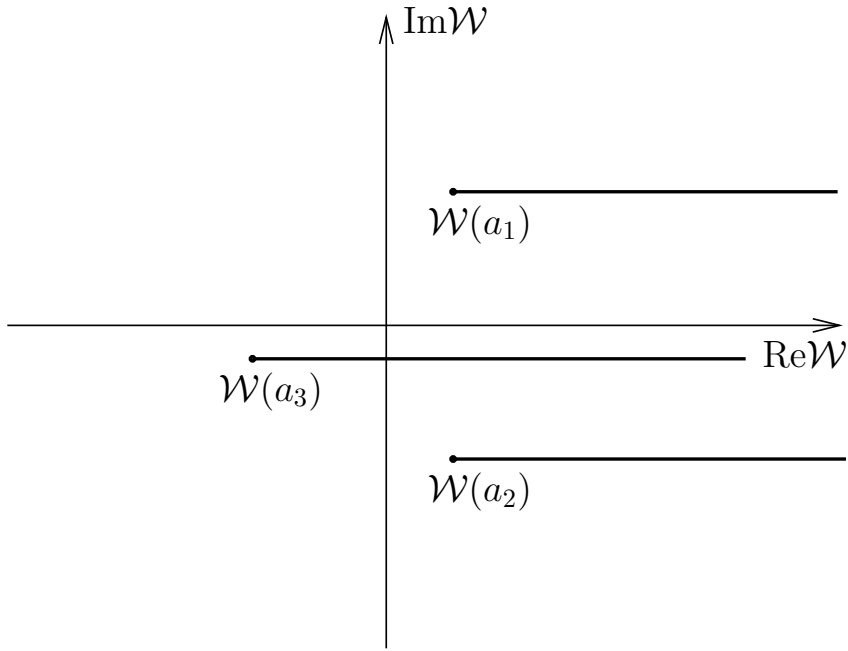}
\end{center}
\caption{Vanishing cycles of the massive $\F^4$ 
LG theory in the ${\cal W}$ plane.}
\label{fig-cycle2}
\end{figure}

Another condition for the boundary curves $\g$ is that they should
start and end at $t\pm\infty$ into a stationary point $\phi=a$ of 
${\cal W}$: these instantonic configurations give rise to the correct
value of the Witten index \cite{vafa}. 
As a consequence, the (A) boundaries can be depicted in the ${\cal W}$
plane as straight half-lines parallel to the real axis, 
that start from a stationary point ${\cal W}(a)$ and go to 
$+\infty$ not crossing any other stationary point by assumption (Figure
\ref{fig-cycle2}):
\be
\g_a\ :\qquad \I{\cal W}(\phi)\ =\ \I{\cal W}(a) \ , \qquad\qquad 
\R{\cal W}(\phi)\ > \ \R{\cal W}(a)\ .
\label{cycle}
\ee
Since a soliton connects two stationary points, these curves are
topologically equivalent to half-solitons and are called vanishing cycles
\cite{cecotti}: the intersection number of two vanishing cycles counts
the number (with sign) of solitons that can extend between the corresponding
stationary points and the Picard-Lefshetz theory describes how
this number changes under monodromy transformations of the cycles
in the complex ${\cal W}$ plane \cite{vafa}.

\begin{figure}[ht]
\begin{center}
\input{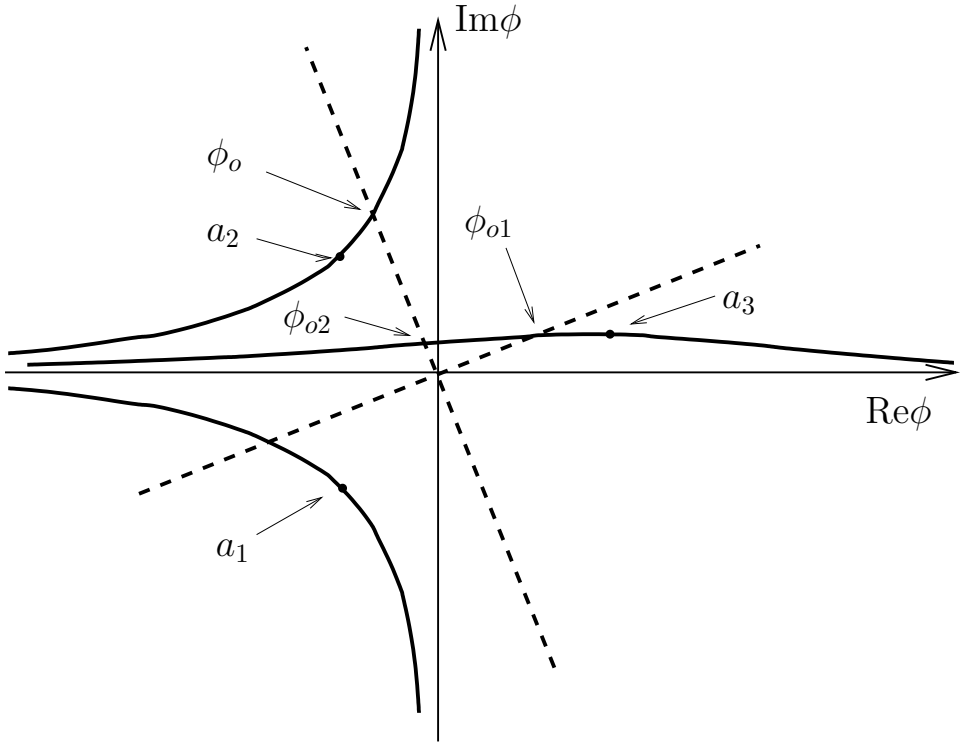}
\end{center}
\caption{Vanishing cycles in the $\phi$ plane for
the LG theory with superpotential ${\cal W}= \F^4/4-\m\F$.
Each cycle contains one stationary point $a_i$
of ${\cal W}$ and intersects the dotted lines either once, $\phi_o$,
or twice, $\phi_{o1},\phi_{o2}$.}
\label{fig-cycle}
\end{figure}


\subsection{Landau-Ginzburg descriptions of A-type 
superconformal boundary conditions}

In order to visualize the D1 brane in field space, i.e. the vanishing
cycle, we should invert the relation between ${\cal W}$ and $\phi$ in
(\ref{cycle}). Let us consider the theory with non-critical potential
${\cal W}=\F^4/4-\m\F$ for definiteness: at $\R{\cal W}=+\infty$, there are
four solutions corresponding to the rays $\phi=\r\exp(ik\pi/2)$,
$k=0,\dots, 3$; from these asymptotes, four curves come to finite
values of $\phi$ (Figure \ref{fig-cycle}).
On the other hand, around one stationary point, say $a_1$, the relation is
quadratic, 
$\R({\cal W}(\phi)-{\cal W}(a_1))\sim\R({\cal W}''\ (\phi-a_1)^2/2)$.
Therefore, the vanishing cycle is the curve made by the two
branches starting from $\phi=a_1$ and approaching two of the four asymptotes;
 it is characterized by the pair of numbers
$(n_1,n_2)$ mod $k+2$ labelling its asymptotes ($n_1\neq n_2$).
Two further vanishing cycles correspond to the other stationary points
of ${\cal W}$ in Figure \ref{fig-cycle}.

We may consider a general massive deformation of the $A_{k+1}$ model, 
${\cal W}=\F^{k+2}+\cdots$, showing $k+1$ distinct stationary points,
$\{a_1,\dots,a_{k+1}\}$; if 
$\R{\cal W}(a_1)<\cdots<\R{\cal W}(a_{k+1})$,
their vanishing cycles will not intersect in the ${\cal W}$ plane 
and thus they will not do either in the $\phi$ plane.
Therefore, this massive theory possesses $k+1$ (A)-type boundaries,
each one characterized by a pair of asymptote numbers.
In the bulk critical limit, e.g. $\m\to 0$,
all these boundary conditions remain present, their curves stick to
the asymptotes and become non-smooth at $\phi=0$. 
Moreover, different massive deformations may produce different set of 
vanishing cycles and their union produces all the $(k+2)(k+1)$ 
boundaries that exist in the superconformal theory, being all the 
pairs of distinct $k+2$ asymptotes.

This is the LG description of superconformal boundary conditions 
found by Hori et al.
\cite{vafa}: indeed, the curves have lost any characterization of the
original massive theory and are just labelled by the universal data
of  the asymptote numbers $(n_1,n_2)$.
Since the $k+2$ asymptotes match one-to-one the vertices of the 
regular polygon drawn before, the numbers $(n_1,n_2)$ can be identified
with the same labels used in (\ref{polyg}) to represent 
the CFT boundaries in the Ramond sector.
In this framework, Hori et al. could identify the
Witten index, namely the overlap of two RR boundary states, 
with the intersection number of the corresponding
vanishing cycles, and could also compute other CFT quantities 
from path-integral expressions in the LG theory.

We now present a slightly improved description of the
A-type boundaries by extending the N=0 approach of 
Section 2 and 3. As shown in the Refs.\cite{maxim}, 
a analytic, polynomial boundary potential $V_b(\phi)$ can be consistently
added to the supersymmetric action (\ref{susy-act}):
\be
S \ =\ S_{\rm bulk~ N=2} +\int_\g \ dt\ \I\left(V_b( \phi) \right)\ .
\label{susy-vb}
\ee
The supersymmetric variation of the action (\ref{susy-max}) with
A-type  boundary conditions now implies a condition
 on the sum of the bulk and boundary potentials at the boundary:
\be
\g\ :\quad \I \wt{\cal W}(\phi) = {\rm const.}\ ,\quad 
\wt{\cal W}\equiv {\cal W} +V_b\ 
\qquad \leftrightarrow\qquad 
\de_t\phi =\pm\ \frac{\de\ov{\wt{\cal W}}}{\de\ov{\phi}} \ .
\label{vb-curve}
\ee
We can take the critical limit in the bulk, 
${\cal W}\to {\cal W}_{\rm crit}=\F^{k+2}/(k+2)$ 
and consider tunable boundary potentials of the form
$V_b(\phi)=a_k\ \phi^{k}+\cdots+ a_1\phi+a_0$, i.e. the deformations
of the Arnold singularity, as done in Section 2. 
All the previous analysis on vanishing cycles
and boundary conditions remains valid upon replacing the non-critical
${\cal W}(\phi)$ with 
$\wt{\cal W}(\phi)={\cal W}_{\rm crit}(\phi) +V_b(\phi)$,
because the two polynomials ${\cal W}$ and $\wt{\cal W}$ are of the same type. 
Furthermore, by choosing different forms of $V_b$ we can realize
all the smooth vanishing cycles and obtain a non-singular
 description of the supersymmetric boundary conditions 
at bulk criticality.
As in the N=0 case, these boundary conditions are not manifestly
conformal invariant, but they became so in the scaling limit
$|\phi|_\g\to\infty$; their universal information is given by the
data of the curves that survives the limit, i.e. by the asymptote labels 
$(n_1,n_2)$.

Next we study the ground state field configurations that are
allowed in the bulk critical theory with boundary potential (\ref{susy-vb}).
The classical equations of motion yield the bulk soliton equations
and the following boundary term:
\ba
0&=& \de_x\phi\ \mp\ i\ \frac{\de\ov{\cal W}_{\rm crit}}{\de \ov{\phi}} \ ,
\label{susy-sol}\\
0&=& \R\left[\d\ov{\phi}\left(\de_x \phi\ 
-\ i\ \frac{\de\ov{V}_b}{\de \ov{\phi}}\right)\right]_\g\ .
\label{susy-eq}
\ea
The fermionic conditions can be obtained from these equations by appropriate
supersymmetric transformations; note also that  
the Cauchy-Riemann equations were used for $V_b$ in the second equation.
The boundary variation should be taken parallel to the $\g$ curve of
the D1 branes, $\d\ov{\phi} \propto\de_t\ov{\phi}|_\g$, whose 
parametrization satisfies (\ref{vb-curve}).
Thus we can rewrite (\ref{susy-eq}) as follows:
\be
0=\R\left[\pm\frac{\de\left({\cal W}_{\rm crit}+V_b\right)}{\de \phi}
\left(\pm\ i\ \frac{\de\ov{\cal W}_{\rm crit}}{\de \ov{\phi}} 
- \ i\ \frac{\de\ov{V}_b}{\de \ov{\phi}} \right)\right]_\g \ ;
\label{susy-check}
\ee
this equation is satisfied by properly choosing the orientation of  
the bulk soliton.

In summary, the boundary potential allows non-trivial  ground state
field configurations at bulk criticality that are static half solitons
satisfying the (A)-type boundary conditions; on the other hand,
the boundary conditions for {\it any} field configuration
are specified by the $\g$ curves in the $\phi$ plane, that 
are parametrized by the other, orthogonal solitons evolving in time.
The bulk and boundary soliton curves intersect both in the $\phi$ and
${\cal W}$ planes; in the $\phi$ plane, 
the bulk solitons (\ref{susy-sol}) are straight lines 
stemming from $\phi=0$:
\be
\R\left[ \phi^{k+2}\right]\ =\ 0\ ,\ \ 
\I\left[ \phi^{k+2}\right]\ >\ 0\ ,\ \leftrightarrow\ 
{\rm Arg}(\phi)= \frac{(2j+1)\pi}{2(k+2)} \ ,j=0,2,\dots, 
\label{susy-ray}
\ee
namely they are rays from the origin that interpose themselves between
the asymptotes of the vanishing cycles (the dotted lines in
Figure \ref{fig-cycle}).
One sees that a ``short'' cycle, $|n_1-n_2|=1$, $\ell=0$,
intersects once the rays of the bulk solitons at the point $\phi_o$,
resulting into a single ground state profile; the longer cycle, 
$|n_1-n_2|=2$ has two intersections, i.e. two ground state profiles,
respectively ending at $\phi_{o1}$ and $\phi_{o2}$.

In the general case, the stationary conditions for
the classical action with  $(\ell,m,s)$ boundary state
yield $\ell+1$ ground state profiles:
contrary to the N=0 case, these multiple stationary points should not
be thought of as representatives of independent boundary conditions,
since they lay on the same D1 brane $\g$, and the free energy should 
be obtained by summing over them,
\ba
&&Z = {\rm e}^{-S_{\rm cl}\left(\phi_{o1}\right)}\ +\cdots
+\ {\rm e}^{-S_{\rm cl}\left(\phi_{o(\ell+1)}\right)}
=\left( \ell+1\right)\ 
{\rm e}^{-S_{\rm cl}\left(\phi_{o1}\right)}\ ,\nl
&&S_{\rm cl}\left(\phi_{oi}\right)\ =\ T\ 
\I\left({\cal W}\left(\phi_{oi}\right)\ + V_b\left(\phi_{oi}\right)\right)\ .
\label{deg-gs}
\ea
Note that the contributions are all equal because the points $\phi_{oi}$
lay on the same vanishing cycle (\ref{vb-curve}).
Moreover, the ground state energy, $E=S_{\rm cl}/T$, 
can be shifted to zero by using the constant
term in $V_b$, thus showing that supersymmetry is preserved.
As described in Section 3, individual stationary points are not
stable semiclassically, unless they are separated by infinite
potential barriers that can be produced in the limit of some parameters
of $V_b$ to infinity.
In the N=2 case, we cannot single out one point out of $\ell+1$, 
say $\phi_{o1}$,
and send the others to infinity without deforming the cycle $\g$
such that its asymptotes are effectively changed, resulting into
another boundary condition $\g'$.
Let us finally note that the variation of the parameters in $V_b$ displaces
the stationary points $\wt{\cal W}(a_i)$ and the corresponding
vanishing cycles change according to the Picard-Lefshetz theory
(Figure \ref{fig-cycle2}) \cite{vafa}.


\subsection{Remarks on the supersymmetric renormalization group flows}

The LG description of the N=2 boundary states presented here
is consistent with the analysis of the N=0 case and it improves the earlier
study of Hori et al. by providing a smooth description at bulk criticality.
On the other hand, the N=2 and N=0 cases present some 
differences: for N=2, we do not see the parametric fine-tuning of $V_b$
that indicates the RG instability of boundary conditions.
Actually, the Cardy-like branes $(\ell,m,s)$ 
considered so far are all stable under the renormalization group
flows preserving supersymmetry, because
they carry no relevant boundary field (besides the identity)
\cite{mms}.

In the Refs.\cite{cosetflows}, some examples of N=2 RG flows were
discussed that start from superpositions of Cardy branes, e.g.
in the $(n_1,n_2)$ notation of the polygon,
\be
(n,n+1)\oplus(n+1,n+2) \ \longrightarrow\ (n,n+2),
\label{susy-flow}
\ee
and similar cases where consecutive sides of the $(k+2)$ polygon are 
added as vectors of the plane to form a single chord.
These flows are due to the extra identity fields that occur
in the superposition of Cardy states; actually, the identity field carries
vanishing $U(1)$ charge and does not break supersymmetry.

Furthermore, the supersymmetric flows are subjected to the 
selection rules of the conservation of the K-theory charges forming
a  $\Z^{k+1}$ lattice \cite{mms}: these charges 
are determined by computing the matrix of Witten indices $I(a,b)$ for
all pairs of RR boundaries $(a)$ and $(b)$; the $(k+1)$-dimensional vector of
charges is assigned to each elementary Cardy branes by analyzing the sub-matrix
of maximal rank $k+1$.

It would be interesting to find the LG descriptions of the
RG flows and the charge assignments, but we cannot presently provide them.
We remark that in terms of the classical free energy, 
the superposition of boundary states at the UV point
in Eq.(\ref{susy-flow}) is not qualitatively different from the IR
boundary, because both amount to superpositions of pairs of ground
state solitons.
However, the ones in the UV pair have different energies, while
those in the IR pair have it equal (cf. (\ref{deg-gs})).
We hope that further analysis will reveal
the parametric instability that is characteristic of RG flows in
the LG approach.


\section{Conclusions}

In this paper, we studied the Landau-Ginzburg theory of 
two-dimensional systems with boundary and obtained a rather simple
picture of the RG flows between the boundary states of
the Virasoro minimal model. This description predicts some selection rules for
the flows that would be interesting to check numerically.
Several related issues would be worth pursuing:
the case with two different boundaries in the strip geometry,
the description of boundary fields and fusion rules,
the study of defects, or kinks \cite{defects}, that allow partial
reflection and transmission and corresponds
to doubles of the boundary states discussed here 
(i.e. to full solitons).
Finally, the boundary entropy $g$ could be computed in this approach
from the semiclassical quadratic fluctuations around the LG 
ground state in the geometry of the disc.

\bigskip
{\large \bf Acknowledgments}

We would like to thank J. Cardy, S. Fredenhagen, K. Graham, M. Henkel,
A. Ludwig, P. Pierce, A. Sagnotti and 
V. Schomerus for interesting discussions and comments.
A. Cappelli thanks the SPhT, Saclay, and LPTHE, Jussieu, for hospitality.
G. D'Appollonio has been supported by the EC Marie Curie postdoctoral
fellowship contract HMPF-CT-2002-01908. 
This work was partially funded by the EC Network contract 
HPRN-CT-2002-00325, 
``Integrable models and applications: from strings to condensed matter''.

\appendix


\section{LG description of the Potts model boundaries}

The partition function on the torus of the three-state Potts model,
\be
Z\ = \ \left\vert\chi_I +\chi_{\eps''}\right\vert^2\ 
    +\ \left\vert\chi_\eps +\chi_{\eps'}\right\vert^2\ 
   +\ \left\vert\chi_{\s_1}\right\vert^2\ 
    +\ \left\vert\chi_{\s_2}\right\vert^2\ 
    +\ \left\vert\chi_{\psi_1}\right\vert^2\ 
    +\ \left\vert\chi_{\psi_2}\right\vert^2\ .
\label{z-pot}
\ee
is diagonal in terms of the six character of the extended chiral algebra 
${\cal W}_3$ and non-diagonal w.r.t. the Virasoro characters, showing
eight left-right symmetric terms.
Therefore, the conformal boundary conditions are expressed in terms of 
six Cardy states, which also respect
the ${\cal W}_3$ symmetry, and two boundaries 
that are just conformal invariant \cite{cardy}\cite{potts}.
The Cardy states form two triplets: they correspond to the 
fixed boundary conditions, $(A), (B), (C)$, that are stable,  
and to the mixed condition $(AB), (BC), (AC)$, that are once unstable.
The corresponding boundary partition functions are \cite{zuber}:
\ba
&&Z_{(A)|(A)} = Z_{(B)|(B)} = Z_{(C)|(C)} = \ \chi_I \ ,\nl
&& Z_{(AB)|(AB)} = Z_{(BC)|(BC)} = Z_{(AC)|(AC)} = \  \chi_I +\ \chi_\eps
\  .
\label{z-trip}
\ea
The two other states are $\Z_3$ singlets that are respectively
called $(f)$ and $(n)$ for free and ``new'' boundary conditions:
they are twice and five-fold unstable, respectively,
\ba
&&Z_{(f)|(f)} = \ \chi_I \ +\chi_{\psi_1}\ +\chi_{\psi_2}\ ,\nl
&& Z_{(n)|(n)} =\  \chi_I +\ \chi_\eps\ +\chi_{\psi_1}\ +\chi_{\psi_2}\
+ \chi_{\s_1}\ +\chi_{\s_2}\  .
\label{z-sing}
\ea
Several boundary RG flows have been found in Ref. \cite{potts}:
for example, the $Z_3$ breaking flow induced by the boundary magnetic field,
\be
(f) \ \longrightarrow\ (AB) \ \longrightarrow\ (A) \ , 
\label{h-flow}
\ee
and the $\Z_3$-preserving flow induced by $\eps$:
\be
(n) \ \longrightarrow\ (f)\ .
\label{eps-flow}
\ee
Other RG flows have been obtained by exploiting the self-duality of the model.

The Landau-Ginzburg description is based on a scalar field with
two real components and again a boundary potential:
\be
S=\int_{x>0} d^2 x \left( \frac{1}{2}\sum_{i=1}^2\ 
\left( \partial_\mu \f_i \right)^2
+ V\left(\f_i\right)\right) \ + \ \int dt\ V_b\left(\f_{oi}\right)\ ,\qquad
\f_{oi}\equiv \f_i(x=0)\ .
\label{LG-pot}
\ee
The stationary conditions are:
\ba
0&=&\frac{\partial^2\f_i}{\partial x^2} - \frac{\partial V}{\partial\f_i} \ ,
\qquad\qquad i=1,2\ ,
\label{eq-pot}\\
0&=&\d\f_{oi}\left(\left.
\frac{\partial\f_i}{\partial x}\right\vert_0 - 
\frac{\partial V_b}{\partial\f_{oi}} 
\right)\ .
\label{bc-pot}
\ea
The boundary conditions on
the two-dimensional field plane may correspond to D0,
D1 and D2 branes and the boundary fields $\f_{oi}$ should be varied
accordingly:
as in the one-dimensional case, we will consider the maximal case
of D2 branes, $\d \f_{oi}\neq 0$,
hoping to find the other cases from localization by the boundary potential.
The equation of motions can be integrated once leading to
the conservation of energy for a ``particle''
starting from one asymptote in the bulk and reaching the boundary with
velocity specified by $V_b$:
\be
\sum_{i=1}^2\ \left(\left. \frac{\de \f_i}{\de x} \right\vert_0 \right)^2
= 2\ V\left(\f_{oi}\right) \ - 2\ V\left(\f_i(\infty)\right) \
=\sum_{i=1}^2\ \left(\frac{\de V_b}{\de \f_{oi}}\right)^2 \ .
\label{en-pot}
\ee
Upon inserting the standard LG
cubic bulk potential (\ref{v-pot}), $V_{\rm crit}=\l\ \R (\f_1+i\f_2)^3$, 
one finds that this equation cannot be discussed in terms of
deformations of the Arnold singularity $D_4$, because 
the corresponding polynomial appears inside a square root. 

Therefore we shall modify the bulk potential into the N=1
supersymmetric form:
\be
V\ = \frac{1}{2}\sum_{i=1}^2 \left(\frac{\de \wt{V}}{\de\f_i}\right)^2\ ,
\qquad \wt{V}=\l\ \R (\f_1+i\f_2)^3 + {\rm deformations}\ .
\label{new-v}
\ee
At the same time, we shall restrict the solutions of the equations of
motion (\ref{eq-pot}) to the solitons,
\be
\frac{\de \f_i}{\partial x} = \pm\ \frac{\partial \wt{V}}{\partial\f_i}\ ,
\label{sol-pot}
\ee
such that the deformations of the bulk potential, the set of
relevant fields and the $\Z_3$ symmetry are the same 
in the two descriptions.

We can now analyze the equations (\ref{bc-pot}) and (\ref{sol-pot}) 
along the same steps of Section 2, finding a relation with
the deformations of the Arnold singularity and discussing the new features
of the two-dimensional problem.  
We consider bulk criticality, 
$\wt{V}= \wt{V}_{\rm crit}= \R (\f_1+i\f_2)^3/3$,
and integrate the soliton equations
(\ref{sol-pot}). They admit the first integral (see Section 4):
\be
\I (\f)^3\ =\ 0\ ,\qquad\qquad \f\equiv \f_1+i\f_2 =\r\ {\rm e}^{i\a}\ , 
\label{ray}
\ee
namely the solitons describe rays in field space emanating from the origin 
at angles $\a=\pi k/3$, $k=0,1,\dots,5$.
Furthermore, the boundary conditions (\ref{bc-pot})
combined with the soliton equations (\ref{sol-pot})
evaluated at the boundary yield an algebraic equation for the
boundary field values that can be written again as the
stationary conditions for the function:
\be
0=\frac{\de W}{\de \f_i}\ , \qquad
W=\mp\wt{V}_{\rm crit} +V_b \ .
\label{w-pot}
\ee
The sign is fixed by requiring that the modulus of the soliton solution
decreases inside the bulk:
\be
0> \frac{\de}{\de x}\left( \f_1^2 +\f_2^2 \right) =
\pm \sum_i\f_i \frac{\de \wt{V}_{\rm crit}}{\de \f_i} =
\pm \R \left(\f_1+i\f_2 \right)^3\ .
\ee
The result depends on the ray spanned by the soliton solution:
\ba
W&=&+\wt{V}_{\rm crit} +V_b\ , \qquad {\rm on}\qquad
\f=\r\ {\rm e}^{i 2 k\pi/3}\ , \qquad k=0,1,2, \nl
W&=&-\wt{V}_{\rm crit} +V_b\ , \qquad {\rm on}\qquad
\f=\r\ {\rm e}^{i (2k+1)\pi/3}\ .
\label{sign-pot}
\ea
The classical action evaluated on the soliton solution takes the
familiar form (cf. (\ref{sol-act})):
\be
S\ =\ T\left( \left\vert\wt{V}_{\rm crit}\right\vert \ + \ V_b \right)\ .
\ee

\begin{figure}[ht]
\begin{center}
\includegraphics[width=8cm]{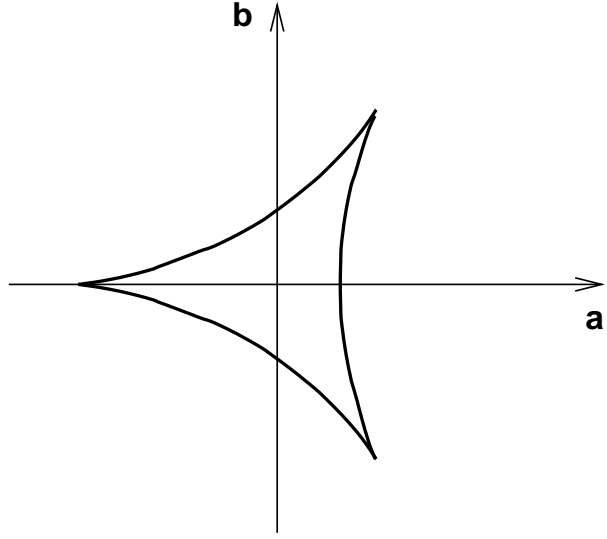}
\end{center}
\caption{Section $c={\rm const.}$ of the parameter space of the $D_4$
singularity.}
\label{fig-cycloid}
\end{figure}

The stationary problem (\ref{w-pot}) involves the study of the
deformations of the singularity $D_4$,
\be
W\ = \ x^3\ -\ 3\ x\ y^2 \ +\ a\ x\ +\ b\ y\ 
+\ c\left(x^2 \ +\ y^2\right) \ ;
\label{d-sing-a}
\ee
we shall first discuss the solutions of this problem and later impose
the restrictions to the rays (\ref{sign-pot}).
The pattern of stationary points can be described in 
the parameter space $(a,b)$ at fixed $c$ values (Figure \ref{fig-cycloid}).
Let us first discuss the $\Z_3$ preserving deformation along the 
$c$ axis: the three-fold singularity at
the origin splits for $c\neq 0$ into a singlet ($x=y=0$)
and a triplet ($|x+i y|=2|c|$) of non-degenerate stationary points,
that can be identified with the four nodes of the $D_4$ Dynkin diagram
($D_4$ is made by a central node connected to each of the three other nodes).
For small $a,b\neq 0$ these solutions slightly move but
remain non-degenerate; when the central node meets one of the three
others, one eigenvalue 
$\l_\pm$ of the Hessian of the stationary point $d W=0$ vanishes.
In complex notations, $z\equiv x+i y$, $\ \eta\equiv a+i b$, these 
conditions are:
\ba
d\ W =0 & \longrightarrow & -\ov{\eta} = \ov{z}^2 + 2\ c\ z\ ;\nl
\l_+ \ {\rm or}\ \l_- = 0 & \longrightarrow & |z| = |c| \ .
\label{eq-cycl}
\ea
Upon replacing $z=c\exp(i\a)$, say for $c>0$, one obtains a curve
in the $(a,b)$ plane in parametric form, $(a(\a),b(\a))$ 
corresponding to a cycloid (Figure \ref{fig-cycloid}):
at the three cusps $\eta=-3c^2\exp(i2n\pi/3)$, $n=0,1,2$, one finds
a twice degenerate stationary point obtained by merging the central
point with two other points, again in agreement with the form of
a three-node sub-diagram of the Dynkin diagram.
For $c<0$ the same curve is obtained with a shift of the parameter
$\a\to\a +\pi$.
Two solutions exist in the region outside the cycloid.

In conclusion, the pattern of deformations of the $D_4$ singularity
is again described by the sub-diagrams of the associated
Dynkin diagram, and is summarized in the following table:
\be
\begin{array}{c|c|c|c}
{\rm degeneracy} & {\rm type} & {\rm Potts\ name} & {\rm deformations} \\
LG\ \ \  CFT & & & LG\ \ \ \ CFT \\
\hline
3 \ \ \ 5 & {\rm singlet} & (n)& 
                (c)\sim\eps;(a,b)\sim(\s_1,\s_2);\psi_1,\psi_2 \\
2\ \ \  / & {\rm triplet} &  /  & / \\
1\ \ \  1 & {\rm triplet} & (AB), (AC), (BC) & (a+b)\sim\eps  \\
0\ \ \  2 & {\rm singlet} & (f) & \psi_1,\psi_2  \\
0\ \ \  0 & {\rm triplet} & (A), (B), (C) & /
\end{array}
\label{LGtable}
\ee
In this table, we also identify the stationary
points of $W$ with the boundary states
of the Potts model according to the following discussion.

The conditions coming from Eq. (\ref{sign-pot}) restrict
the solutions found before to lay on one of the rays from the origin
at angles $k\pi/3$, for example on $\f =\r >0$:
a pair of solutions can be found on that ray for $c<0$ and specific values of
$(a,b)$, that merge at one point into a once-degenerate solution;
however, the simultaneous merging of three points cannot be realized
for the parameters at one cusp of the cycloid, because two solutions
would be coming from outside the ray. 
Otherwise said, two directions out of the twice degenerate points
are frozen by the conditions (\ref{sign-pot}), and thus these points
are effectively ordinary stable points.
As for the detuning of the highest singularity at $\f=0$, the restriction of
the solutions on the rays put conditions on the phase of $a+i b$, such that
the one out of the two real parameters is actually discrete.

After imposing these conditions, the LG description of the boundary states
of the Potts model is rather satisfactory, see Table (\ref{LGtable}); 
besides the $Z_3$ symmetry, the identification is based on the
sets of deformations as given by operator content, Eqs. 
(\ref{z-trip},\ref{z-sing}), on one side, and by the parameters
$(a,b,c)$ on the other side, keeping in mind that the deformations 
corresponding to the $\psi_i$ fields are missing in the LG description, 
as explained in Section 3.3.
Note that the RG flows of the Potts model,
\be
(n) \ \longrightarrow \ (AB) \ \longrightarrow\ (A) \ , \qquad
(n) \ \longrightarrow \ (f)\ ,
\ee
are reproduced, but the flow (\ref{h-flow}) is not, because
$(f)$ is stable in the LG description.

In conclusion, it seems that the $\psi_i\to 0$ projection of the space of
boundary RG flows of the Potts model is reproduced: however,
our analysis has been rather limited and has relied 
on a number of additional hypotheses.


\def\NPB#1#2#3{{\it Nucl.~Phys.} {\bf{B#1}} (#2) #3}
\def\CMP#1#2#3{{\it Commun.~Math.~Phys.} {\bf{#1}} (#2) #3}
\def\CQG#1#2#3{{\it Class.~Quantum~Grav.} {\bf{#1}} (#2) #3}
\def\PLB#1#2#3{{\it Phys.~Lett.} {\bf{B#1}} (#2) #3}
\def\PRD#1#2#3{{\it Phys.~Rev.} {\bf{D#1}} (#2) #3}
\def\PRL#1#2#3{{\it Phys.~Rev.~Lett.} {\bf{#1}} (#2) #3}
\def\ZPC#1#2#3{{\it Z.~Phys.} {\bf C#1} (#2) #3}
\def\PTP#1#2#3{{\it Prog.~Theor.~Phys.} {\bf#1}  (#2) #3}
\def\MPLA#1#2#3{{\it Mod.~Phys.~Lett.} {\bf#1} (#2) #3}
\def\PR#1#2#3{{\it Phys.~Rep.} {\bf#1} (#2) #3}
\def\AP#1#2#3{{\it Ann.~Phys.} {\bf#1} (#2) #3}
\def\RMP#1#2#3{{\it Rev.~Mod.~Phys.} {\bf#1} (#2) #3}
\def\HPA#1#2#3{{\it Helv.~Phys.~Acta} {\bf#1} (#2) #3}
\def\JETPL#1#2#3{{\it JETP~Lett.} {\bf#1} (#2) #3}
\def\JHEP#1#2#3{{\it JHEP} {\bf#1} (#2) #3}
\def\TH#1{{\tt hep-th/#1}}

\end{document}